\documentclass[twocolumn,
 amsmath,amssymb,
 aps,
 pra,
]{revtex4-1}
\usepackage{graphicx}
\usepackage{bm}      
\usepackage{esint}
\usepackage{pdfpages}
\usepackage{etoolbox} 

\makeatletter
\patchcmd{\@outputpage@head}{\@ifx{\LS@rot\@undefined}{}{\LS@rot}}{}{}{}
\makeatother

\newcommand*{\bra}[1]{\left\langle #1 \right\vert}
\newcommand*{\ket}[1]{\left\vert #1 \right\rangle}

\newcommand*{\dotprod}[2]{\left\langle #1 \middle| #2 \right\rangle}
\newcommand*{\expect}[3]{\left\langle #1 \middle| #2 \middle| #3 \right\rangle}
\newcommand*{\matr}[1]{\ensuremath{\underline{\underline{#1}}\,}}

\newcommand*{\im}[1]{\mathrm{Im}\left\lbrace #1\right\rbrace}

\newcommand*{\tr}[2]{\mathrm{Tr}_{#1}\left\lbrace#2\right\rbrace}
\usepackage{color}
\definecolor{blue}{RGB}{31,119,180}
\definecolor{green}{RGB}{44,160,44}
\definecolor{orange}{RGB}{255,127,14}

\usepackage[normalem]{ulem}

\begin{document}
\title{Mechanistic Investigations of Electronic Current Dynamics Through a Single-Molecule-Graphene--Nanoribbon Junction}

\author{Vincent Pohl}
\email{v.pohl@fu-berlin.de}
\affiliation{Institute for Chemistry and Biochemistry, Freie Universit\"at Berlin, Takustra\ss{}e 3, 14195 Berlin, Germany}

\author{Lukas Eugen Marsoner Steinkasserer}
\affiliation{Institute for Chemistry and Biochemistry, Freie Universit\"at Berlin, Takustra\ss{}e 3, 14195 Berlin, Germany}

\author{Jean Christophe Tremblay}
\email{jean-christophe.tremblay@univ-lorraine.fr}
\affiliation{Institute for Chemistry and Biochemistry, Freie Universit\"at Berlin, Takustra\ss{}e 3, 14195 Berlin, Germany}
\affiliation{Laboratoire de Physique et Chimie Th\'eoriques CNRS-Universit\'e de Lorraine, UMR 7019, ICPM, 1 Bd Arago, 57070 Metz, France}

\begin{abstract}
To assist the design of novel, highly efficient molecular junctions,
a deep understanding of the precise charge transport mechanisms through these devices is of prime importance. 
In the present contribution, we describe a procedure to investigate spatially-resolved electron transport through a nanojunction, 
at the example of a nitro-sub\-sti\-tu\-ted oligo-(phenylene ethynylene) covalently
bound to graphene nanoribbon leads.
Recently, we demonstrated that the conductivity of this single-molecule-graphene-nanoribbon junction 
can be switched quantitatively and reversibly upon application of a static electric field in a top gate position,
in the spirit of a traditional field effect transistor [\textit{J.~Phys.~Chem.~C}, 2016, \textbf{120}, 28808--28819].
The propensity of the central oligomer unit to align with the external field was found to induce a damped
rotational motion and to cause an interruption of the conjugated $\pi$-system,
thereby drastically reducing the conductance through the nanojunction. 
In the current work, we use the driven Liouville-von-Neumann (DLvN) approach for time-dependent electronic transport calculations
to simulate the electronic current dynamics under time-dependent potential biases for the two logical states of the nanojunction.
Our quantum dynamical simulations rely on a novel localization procedure using an orthonormal set of molecular orbitals obtained from
a ground state density functional theory calculation to generate a localized representation for the different parts of the molecular junction.
The transparent DLvN formalism allows us to directly access the density matrix, and it captures both the non-Markovian scattering dynamics and the relaxation to the stationary limit. 
Using this time-dependent one-electron density matrix, it is possible to reconstruct the time-dependent electronic current density, unraveling insightful time-dependent mechanistic details of the electron transport.
\end{abstract}

\maketitle

\section{Introduction}
The research field of molecular electronics\cite{ratner2013brief_history,Sun2014review_single_mol_electronics,Xiang2015reviewMolecularScaleElectronics} aims at designing electronic components
such as wires\cite{JamesTour2005molWires,Laurens2011molWires},
rectifiers\cite{Aviram1974rectification,Joachim2000rectification,Kornilovitch2002rectification,Liu2006rectification,DezPrez2009rectification,Nijhuis2010rectification,Yee2011rectification,Hihath2011rectification,Wang2014rectification,Trasobares2016rectification,wang2017review_modulation},
LEDs\cite{Recht2014molecularLED,Yang2015molecularLED,Goswami2015molecularLED,Braun2015molecularLED},
or transistors\cite{Dulic2003ring_opening_exp,Pati2004pi_sigma_pi,Li2004ring_opening,Mendes2005redox,Choi2006STM,delValle2007cis_trans,Liljeroth2007tautomerization,Huang2007diaryletheneMetal,Zhao2008chineseCPD_Metal,BeneschThoss2009H_transfer,Pan2009H_transfer,Cai2011diaryletheneGNR,Roldan2013CPD,Pathem2013Dihydroazulene,Wu2014chineseCPDswitch,Xie2015nonbondingGNR,Chen2015DihydroazuleneMetal,Kumagai2015review,RodrguezManzo2016gnr_FET,Xia2016cis_trans,Kumagai2016tautomerism,Xin2017realGrapheneSwitch}, 
as well as whole electronic devices from single molecules.
Especially transistors, which are the central switching units in modern electronics, play a prominent role in this field, 
and many systems have been proposed and investigated, both theoretically and experimentally, 
to bring transistors to molecular scale (see for example Ref.~\citep{Fuentes2011reviewMolecularSwitches,Pathem2013reviewMolecularSwitches,Perrin2015review_molFET,Zhang2015reviewMolecularSwitches} and references therein).
Recently, the first light-induced molecular switch operating reliably at room temperature has been synthesized.\cite{Jia2016realGrapheneSwitch}
In this case, a diarylethene molecule, which is well known for its photoinduced switching behavior,
is bound covalently between two graphene tips.
The covalent nature of the bond renders the nanojunction very stable and allows for an enhanced electronic coupling to the leads, which positively affects the conductivity of the junction.\cite{Jia2015reviewGrapheneContacts}
Graphene-based nanostructures such as carbon nanotubes and graphene nanoribbons (GNRs) appear as very natural choices for contacts and wires, 
because they behave like one-dimensional conductors and exhibit unique electronic transport properties even at room temperature \cite{Schwierz2010graphene_transistors}.

\begin{figure}[tb]
\centering\includegraphics[width=3in]{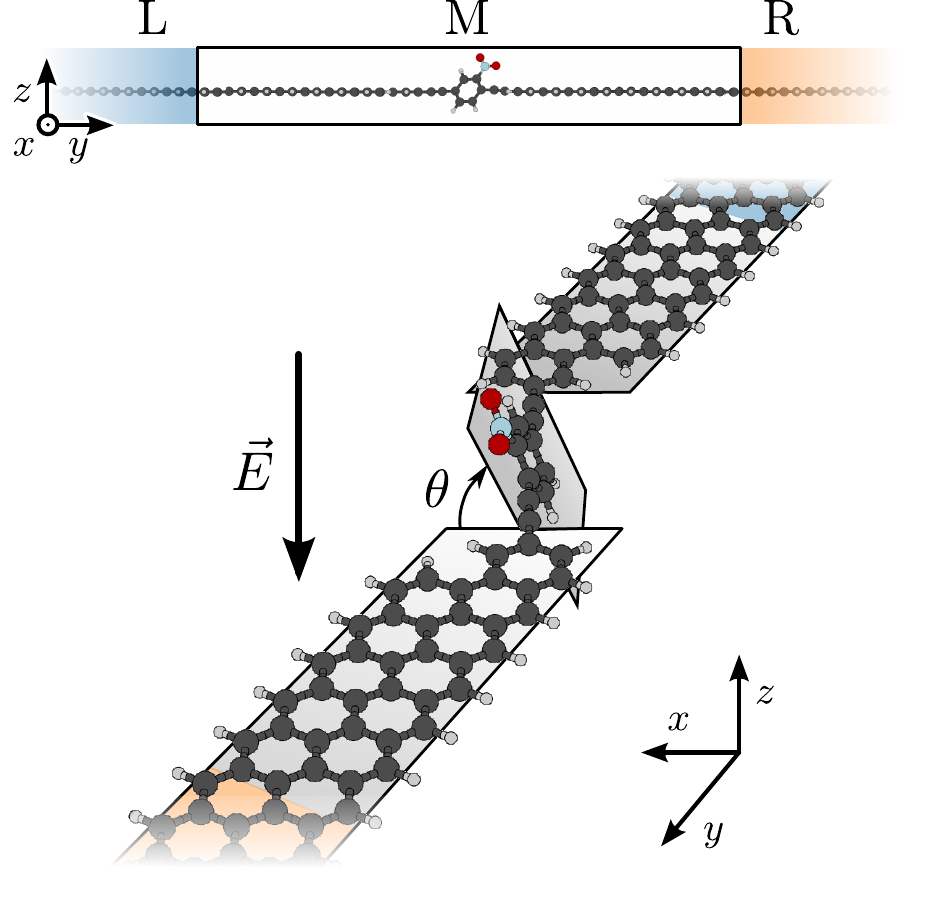}
\caption{\label{fig:switch_sketch}
Cartoon of the investigated OPE-GNR nanojunction viewed from different perspectives.
Carbon atoms are colored in dark grey, hydrogen atoms in light grey, nitrogen atoms in light blue, and oxygen atoms are colored in red.
As demonstrated in Ref.~\citep{PohlTremblay2016opegnr}, 
an electric field along the $z$-axis, $\vec{E}$, can be used to switch the device from a 
conducting state (ON, $\theta\approx 0^\circ$) to a less conducting state (OFF,  $\theta\approx 90^\circ$).
For the investigation of the electron current dynamics, the device is divided into three segments:
the left lead (L, highlighted in blue), the extended molecule (M, black solid box), and the right lead (R, highlighted in orange).
The color code and the coordinate system defined in this figure are used throughout this work.
} 
\end{figure}

Besides diarylethenes, oligo-(phenylene ethynylene)s (OPEs) have attracted great attention especially as
molecular wires (see, e.g., Ref.~\citep{JamesTour2005molWires} and references therein), and they appear as another promising class of molecular switching units.
When attached to gold surfaces, OPEs demonstrated promising switching properties in STM experiments\cite{Donhauser2001florida_exp,Ramachandran2003expOPE,Moore2006expOPE_NO2}.
Although the mechanism observed in these experiments is rather of stochastic nature (for an overview of the proposed mechanisms see Ref.~\citep{Zhang2015reviewMolecularSwitches}), 
a rotation of the central phenyl group about the triple bonds was theoretically proposed as a possible
switching mechanism.
This assumption was the starting point for the investigation of Agapito and Cheng\cite{AgapitoCheng2007florida},
who investigated a nanojunction where a nitro-substituted OPE wire was bound to two 3-{\it zigzag} graphene nanoribbons (ZGNRs) serving as electrodes.
This OPE-GNR model system is sketched in Fig.~\ref{fig:switch_sketch}.
In their work, non-equilibrium Green's function (NEGF) simulations revealed the markedly different behavior of two logical states, one planar with high (ON, $\theta\approx 0$) and one perpendicular conformer with low conductivity (OFF, $\theta\approx 90^{\circ}$). 
It was demonstrated that the rotation of the central nitrophenyl-group about the triple bonds causes a breakdown of the conjugation of the $\pi$-system, leading to a significant drop in conductivity.
Recently, we have proposed a practical procedure to drive this system from the ON to the energetically unfavored OFF conformer dynamically by applying an external static electric field $\vec{E}$, in the spirit of a traditional field effect transistor.\cite{PohlTremblay2016opegnr}
Ground state nuclear quantum dynamics simulations of the complete switching cycle
within the reduced density matrix formalism showed that the system can be reliably switched on and off
without showing any memory effects.

The present contribution focuses on another fundamental dynamical aspect of the same system:
the time-dependent electron transport under non-equilibrium conditions at finite bias voltage.
To this end, we resort to the recently developed driven Liouville-von-Neumann (DLvN) approach for time-dependent electronic transport calculations\cite{Sanchez2006referred_by_Hod,Subotnik2009referred_by_Hod,ZelovichHod2014eTransport,ChenFranco2014simple_accurate_eTransport,ZelovichHod2015real_vs_state_space,Hod2016lindblad_form,ZelovichHod2016nonorthogonal,ZelovichHod2017paramter_free,Morzan2017parameter_free,OzHod2018}.
In conventional fashion, the finite system is first divided into three parts: the left lead (L), the extended molecule (M), and the right lead (R).
The associated equations of motion for the density matrix in this localized basis are supplemented by a driving term which aims at preserving a steady-state subject to non-equilibrium boundary conditions.
Charge transport through the nanojunction arises from imposing, e.g., finite temperatures to the leads 
or applying a potential bias voltage between the electrodes.
In the original formulation of the DLvN approach as well as in the standard NEGF treatment,
localized atom centered basis functions are used to define the pseudo spectral basis functions of different parts of the system.
For realistic systems, this basis set can become very large and basis functions cannot be safely neglected, since they are all coupled to each other.
Therefore in the present work, this approach is modified to reduce its computational effort and improve its scalability towards larger molecular systems.
Here, an orthonormal set of delocalized molecular orbitals is first computed from a single ground-state density functional theory (DFT) calculation for the extended molecule and sections of the leads.
A numerical localization procedure is then used to project a subset of these molecular orbitals at energies close to the Fermi level onto the different sections of the nanojunction.
In order to increase the system size without jeopardizing its computational efficiency and  to converge the results to a reference NEGF calculation, we further describe how to parameterize an effective microscopic tight-binding Hamiltonian in which the leads can be readily extended.

The DLvN approach stands out by its simplicity and transparency, providing direct access to the system's density matrix, which can be used to straightforwardly compute a multitude of observables. 
One of which, the time-dependent (local) electronic current density (also called ``electronic flux density''),
is the main focus of this work.
This vector field provides a spatially resolved picture of the instantaneous flow of electrons and
allows for an intuitive interpretation of the electron dynamics, revealing the details of the electron transport mechanism.
The precise knowledge of the electron flow mechanism is heralded as the key aspect for the design of more efficient molecular junctions.\cite{Frisbie2016Comment_on_Jia2016}
Despite these promises, example applications and investigations of the electronic current density in molecular junctions remain few and far apart.\cite{SaiBushong2007EFD_junction,WenChen2011CurrentDensity,Paez2015EFD_GNRdefect,Walz2015EFD_GNR,He2016EFD_ZGNR,Nozaki2017efdJunction}
For one, the underlying electronic structure are often based on parametric tight-binding (TB)  methods\cite{Ernzerhof2006EFD_jucntion,Solomon2010EFD_jucntion,RaiHod2010circularEFD}, 
allowing only for the investigation of steady-state local currents from site to site.
Alternatively, the stationary electronic current density can be extracted from NEGF calculations 
as an incoherent sum over all channels open at a given potential bias \cite{WilhelmWalz2014EFD_GNR,Walz2014EFD_GNR,Walz2015EFD_GNR,Wilhelm2015EFD_GNR,Nozaki2017efdJunction}. 
In this work, we aim at addressing the need for a better mechanistic understanding of electron flow in nanojunctions under non-equilibrium conditions using a simple dynamical formalism based on microscopic characterization of the electronic structure.
This can be achieved using explicitly time-dependent DFT simulations
\cite{He2016EFD_ZGNR,Stefanucci2006TDDFT_transport,ZhengChen2007TDDFT_transport,YamChen2011TDDFT_transport,Schaffhauser2016TDDFT_transport,Chen2018TDDFTOS_transport},
which maps the many-electron density on a time-local one-electron density via the holographic theorem.
We follow here an approach based on the DLvN formalism, which does not rely on a
time-local approximation of the exchange-correlation potential to map the time-dependent many-electron density. 
Consequently, it is possible to describe from a single simulation
the non-Markovian scattering dynamics on the attosecond timescale, 
the equilibration dynamics in the femtosecond regime,
and the stationary limit of the current after a few picoseconds.

The paper is structured as follows. 
Sec. ``Methodology'' introduces the Driven Liouville-von-Neumann equation, the localization scheme,
and the analysis toolset for the electron dynamics.
In Sec. ``Computational Details'', the numerical methods and technical details are described.
The results are presented and analyzed in Sec. ``Results and Discussion'',
before concluding remarks summarize our most important findings.

\section{Methodology\label{Methodology}}
\subsection{Driven Liouville-von-Neumann Equation}

Within the framework of the driven Liouville-von-Neumann (DLvN) approach for time-dependent electronic transport simulations, 
a finite molecular junction is formally divided into three parts: the left lead (L), the extended molecule (M), and the right lead (R) (cf. Fig.~\ref{fig:switch_sketch}).
In the localized representation, the time-evolution of the system is described by\cite{ZelovichHod2016nonorthogonal,ZelovichHod2017paramter_free}
\begin{widetext}
\begin{equation}\begin{aligned}\label{eq:DLvN}
  \frac{\partial \matr{\rho}(t)}{\partial t} &= -\frac{\imath}{\hbar}\left[\matr{H}_\mathrm{sys},\matr{\rho}(t)\right]-\frac{\imath}{\hbar}\left[\imath \matr{W},\matr{\rho}(t)\right]_+ \\
  &= -\frac{\imath}{\hbar}\left[\matr{H}_\mathrm{sys},\matr{\rho}(t)\right] 
  - \frac{1}{2\hbar}  \begin{pmatrix}
	      \Big[\matr{\varGamma}_\mathrm{L},\Big(\matr{\rho}_\mathrm{L}(t)-\matr{\rho}_\mathrm{L}^0 \Big)\Big]_{+} & \matr{\varGamma}_\mathrm{L}\matr{\rho}_\mathrm{LM}(t) & \matr{\varGamma}_\mathrm{L}\matr{\rho}_\mathrm{LR}(t)+\matr{\rho}_\mathrm{LR}(t)\matr{\varGamma}_\mathrm{R} \\
	      \matr{\rho}_\mathrm{ML}(t)\matr{\varGamma}_\mathrm{L} & 0 & \matr{\rho}_\mathrm{MR}(t)\matr{\varGamma}_\mathrm{R} \\
	      \matr{\varGamma}_\mathrm{R}\matr{\rho}_\mathrm{RL}(t)+\matr{\rho}_\mathrm{RL}(t)\matr{\varGamma}_\mathrm{L} & \matr{\varGamma}_\mathrm{R}\matr{\rho}_\mathrm{RM}(t) & \Big[\matr{\varGamma}_\mathrm{R},\Big(\matr{\rho}_\mathrm{R}(t)-\matr{\rho}_\mathrm{R}^0\Big) \Big]_{+}
	    \end{pmatrix},
\end{aligned}\end{equation}
\end{widetext}
where $\hbar$ is the reduced Planck constant, $\matr{\rho}(t)$ is the one-particle reduced density matrix, and $\matr{H}_\mathrm{sys}$ refers to the system's Hamiltonian.
The diagonal matrix $\matr{\rho}_{\ell}^0$ describes the equilibrium Fermi-Dirac statistics
of the respective lead $\ell=\{\mathrm{L, R}\}$,
\begin{equation}\label{eq:fermidirac}\begin{aligned}
  f_{\ell}\big(\varepsilon_a^{\ell}\,\big) = \frac{1}{\exp\left[\left(\varepsilon_a^{\ell}-\mu_{\ell}\right)\middle/k_\mathrm{B}T_{\ell}\right]+1}
\end{aligned}\end{equation}
with the Boltzmann constant $k_\mathrm{B}$, the lead state energies $\varepsilon^{\ell}_a$, the electronic temperature $T_{\ell}$, and the chemical potential $\mu_{\ell}$.
While the first term on the right-hand side of Eq.~\eqref{eq:DLvN} describes the coherent time-evolution
of the system, the second term containing the complex Hamiltonian $\imath \matr{W}$ is driving the system dynamically towards a non-equilibrium situation at a
rate ${\matr{\varGamma}_{\ell}}/{\hbar}$.
This term can be attributed to the coupling of the finite lead section to an implicit semi-infinite electronic reservoir within the wide-band approximation.\cite{ZelovichHod2017paramter_free}
It is often defined as a constant factor that
is either adjusted to the reflection time scales in a finite junction model\cite{ZelovichHod2014eTransport},
or fitted to a NEGF reference calculation\cite{ZelovichHod2015real_vs_state_space,Hod2016lindblad_form,ZelovichHod2016nonorthogonal}. 
Recently, the DLvN approach has been extended to parameter-free state-dependent broadening factors that allow for a more accurate description of the couplings between the finite lead section and the electronic reservoir.\cite{ZelovichHod2017paramter_free}
This is the approach we choose to follow in the present work, and we will review the procedure below.

The influence of a semi-infinite reservoir on a finite lead section is described by the reservoir's
retarded self-energy matrix 
\begin{equation}\begin{aligned}
  &\matr{\Sigma}_\mathrm{res}^r\big(\varepsilon\big) 
  \\&\quad= \Big(\epsilon^+\, \matr{S}_{\ell, \mathrm{res}} - \matr{V}_{\ell, \mathrm{res}}\Big)\matr{G}_\mathrm{res}^{r,0}\big(\epsilon^+\big) \Big(\epsilon^+\, \matr{S}_{\mathrm{res}, \ell} - \matr{V}_{\mathrm{res}, \ell}\Big)\phantom{\Bigg|}
\end{aligned}\end{equation}
where $S_{\ell, \mathrm{res}}$ is the overlap between a lead and the reservoir, $V_{\ell, \mathrm{res}}$ is the corresponding coupling matrix, and $\epsilon^+=\epsilon+\imath\eta$ with $\eta\rightarrow0^+$. 
The retarded surface Green's function of the isolated reservoir is given by
\begin{equation}\begin{aligned}
  \matr{G}_\mathrm{res}^{r,0}\big(\varepsilon\big) = \Big[\epsilon^+\,\matr{S}_\mathrm{res} - \matr{H}^0_\mathrm{res}\Big]^{-1}
\end{aligned}\end{equation}
with the Hamiltonian matrix of the uncoupled semi-infinite reservoir $\matr{H}^0_\mathrm{res}$ and the
overlap matrix $\matr{S}_\mathrm{res}$.
For each lead state $\ket{\varphi^{\ell}_a}$, a dressed Hamiltonian is constructed by adding
the reservoir's self-energy evaluated at the energy of the respective lead eigenenergy, $\matr{\Sigma}_\mathrm{res}^r\big(\varepsilon^{\ell}_a\big)$,  to the lead Hamiltonian $\matr{H}_{\ell}$.
A subsequent diagonalization of this new Hamiltonian matrix, yields new dressed eigenstates and eigenenergies.
By gradually turning on $\matr{\Sigma}_\mathrm{res}^r\big(\varepsilon^{\ell}_a\big)$,
it is possible to follow a given state of the undressed Hamiltonian.
In matrix notation, the dressed eigenvector corresponding to the undressed state $\ket{\varphi^{\ell}_a}$ reads
\begin{equation}\begin{aligned}
  \left(\matr{H}_{\ell}+ \matr{\Sigma}_\mathrm{res}^{r}\!\big(\varepsilon^\mathrm{\ell}_a\big)\right)\matr{U}_{{\ell}_a} 
    = \matr{U}_{{\ell}_a} \matr{\Lambda}_{{\ell}_a},
\end{aligned}\end{equation}
The level broadening $\matr{\varGamma}_{\ell}^{(a,a)}$ caused by the finite life time of this state
is given by the imaginary part of the dressed eigenvalue
\cite{Henderson2006dressedHamiltonian,Liu2014dressedHamiltonian,ZelovichHod2017paramter_free}
\begin{equation}\begin{aligned}\label{eq:broadening}
  \matr{\varGamma}_{\ell}^{(a,a)} = -2\, \im{\matr{\Lambda}_{{\ell}_a}}^{(a,a)}
\end{aligned}\end{equation}

\begin{figure}[tb]
\centering\includegraphics[width=3in]{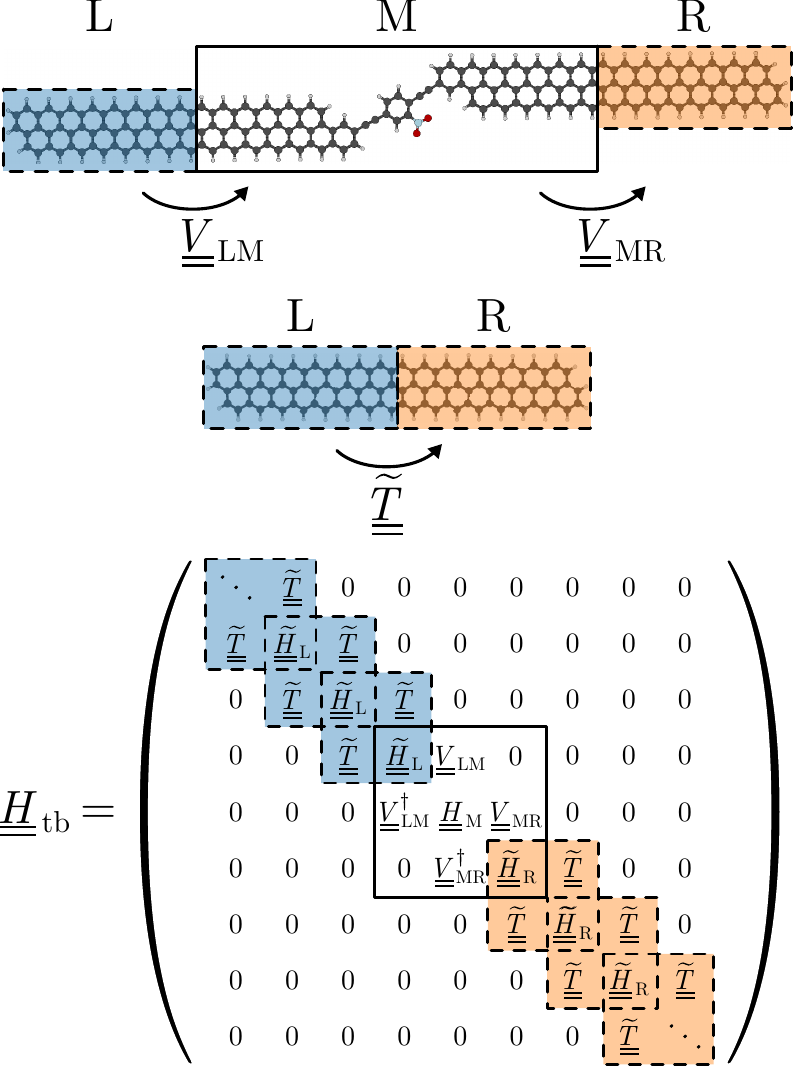}
\caption{\label{fig:concept}
Upper panel: Ball-and-sticks representation of the finite OPE-GNR model system used within the work. 
The nanojunction is divided into three parts: the left lead (L,  in blue), the extended molecule (M, black solid box), and the right lead (R,  in orange).
Central panel: Ball-and-sticks representation of the lead dimer composed of a left lead part (L,  in blue) and a right lead part (R,  in orange). 
Bottom panel: 
Conceptual sketch of the different contributions to the tight-binding Hamiltonian matrix $\matr{H}_\mathrm{tb}$.
This Hamiltonian is parametrized by localizing the molecule's (upper panel) molecular orbitals onto the three parts of the device, and extending the leads using the matrix elements of the dimer (central panel). 
The tilde denotes matrix blocks obtained from the lead dimer calculation.}
\end{figure}

\subsection{Model Construction}

Fig.~\ref{fig:concept} (upper panel) shows a sketch of the finite model system used in this work, with the three parts highlighted as colored areas. 
In contrast to previous studies, the starting point for the localization procedure is an orthonormal set of molecular orbitals (MOs) $\{\ket{\varphi_a}\}_{n_{\mathrm{MO}}}$ 
and their corresponding eigenenergies $\vec{\varepsilon}=\{\varepsilon_a\}_{n_{\mathrm{MO}}}$.
These are obtained from a ground state density functional theory (DFT) calculation, which satisfy a one-electron 
Kohn-Sham equation of the form
\begin{equation}\begin{aligned}\label{eq:H_KS}
  \hat{h}_{\textrm{KS}}\ket{\varphi_a} =  \varepsilon_a \ket{\varphi_a}
\end{aligned}\end{equation}
with the Hamiltonian $\hat{h}_{\textrm{KS}}=-\frac{\hbar^2}{2m_e}\nabla_e^2+\hat{v}_{\textrm{KS}}$, where $m_e$ is the mass of an electron and $\hat{v}_\mathrm{KS}$ is the Kohn-Sham potential operator.
For molecular systems composed of $N_\alpha$ atoms, 
MOs are usually expanded in a finite set of atom-centered orbitals (AOs)
\begin{eqnarray}\label{molcao}
\varphi_{a} \big(\vec{r}\,\big) = \sum_{\alpha=1}^{N_{\alpha}}\sum_{i_\alpha=1}^{n_{\mathrm{AO}}{(\alpha)}} D^{(a)}_{i_\alpha} \chi_{i_\alpha}\big(\vec{r}-\vec{R}_{\alpha}\big),
\end{eqnarray}
where $\vec{r}$ and $\vec{R}_{\alpha}$ are the coordinates of an electron and of nucleus $\alpha$, respectively.
The index $n_{\mathrm{AO}(\alpha)}$ defines the number of AOs, $\chi_{i_\alpha}\big(\vec{r}-\vec{R}_{\alpha}\big)$, centered on atom $\alpha$.

Modern theoretical approaches to electronic transport, such as non-equilibrium Green's functions and the DLvN ansatz, describe the dynamics of electrons in terms of pseudo-spectral states localized on different parts of the nanojunction: two leads, and the scattering region.
Since the MOs obtained from a ground state quantum chemistry calculations are generally delocalized over the whole extent of the nanostructure, localization onto each of the three sections is required.
In the present work, this is achieved by following a sequential procedure involving  numerical unitary
basis set transformations of a selected subset of MOs within a specific energy window.
This bottom-up approach drastically reduces the basis size 
while conserving the orthogonality of the MOs obtained from conventional quantum chemistry calculations.

\subsubsection{Defining Localized Lead States}
To first define a set of localized lead states,
we construct a lead dimer, as depicted in Fig.~\ref{fig:concept} (central panel).
This prevents any artificial influence of the asymmetry of central group on the leads.
The one-electron Kohn-Sham equation of this dimer is given by
\begin{equation}\begin{aligned}
  \hat{h}_{\textrm{KS}}\ket{\widetilde{\varphi}_a} =  \widetilde{\varepsilon}_a \ket{\widetilde{\varphi}_a}.
\end{aligned}\end{equation}
For clarity, all matrices represented in the MO basis of the lead dimer are denoted with a tilde. 
Note that the primitive atomic orbital basis and the relative position of the lead atoms are identical to those of the complete system (cf. Fig.~\ref{fig:concept} (upper panel)).
In order to localize the MOs on the dimer units,
a linear operator quantifying the differential projection on the right and left leads is used to define a linear metric as
\begin{eqnarray}\label{eq:loc_lr_dimer}
  \big(\widetilde{\matr{R}}-\widetilde{\matr{L}}\big)\widetilde{\matr{Q}} = \widetilde{\matr{Q}}\widetilde{\Lambda}_\mathrm{RL}
\end{eqnarray}
where  
\begin{equation}\begin{aligned}
  \widetilde{\matr{L}}_{ab} &= \expect{\widetilde{\varphi}_a}{\hat{{P}}_\mathrm{L}}{\widetilde{\varphi}_b}\\
  &= \sum_{\substack{\alpha_{\phantom{j}}\\\beta\in \{ \mathrm{L}\}}}\sum_{\substack{i_\alpha\\j_\beta}}  \widetilde{D}^{(a)}_{i_\alpha} \widetilde{D}^{(b)}_{j_\beta}\dotprod{\widetilde{\chi}_{i_\alpha}}{\widetilde{\chi}_{j_\beta}},
\end{aligned}\end{equation}
and accordingly
\begin{equation}\begin{aligned}
  \widetilde{\matr{R}}_{ab} 
  = \sum_{\substack{\alpha_{\phantom{j}}\\\beta\in \{ \mathrm{R}\}}}\sum_{\substack{i_\alpha\\j_\beta}}  \widetilde{D}^{(a)}_{i_\alpha} \widetilde{D}^{(b)}_{j_\beta}\dotprod{\widetilde{\chi}_{i_\alpha}}{\widetilde{\chi}_{j_\beta}}.
\end{aligned}\end{equation}
Here, $\hat{{P}}_\mathrm{L/R}$ is the Mulliken projector onto the atoms of the left or right lead, respectively. 
The spectrum of the operator $(\widetilde{\matr{R}}-\widetilde{\matr{L}})$ gives a measure of the localization
of the MOs on the leads.
The eigenvalues $\widetilde{\Lambda}_\mathrm{RL}<0$ correspond to a localization onto the left lead
(blue shaded area in the central panel of Fig.~\ref{fig:concept}), and $\widetilde{\Lambda}_\mathrm{RL}>0$
to a localization onto the right lead (orange shaded area in Fig.~\ref{fig:concept}).
To avoid artificial mixing between energetically widely separated states, an energy weighting, 
$\mathrm{exp}\big(-(\epsilon_a-\epsilon_b)^2/2\sigma^2\big)$ with $\sigma=1\,\mathrm{eV}$,
has been applied to the coefficients in Eq.~\eqref{eq:loc_lr_dimer}.
Subsequent diagonalization of the resulting diagonal blocks of the Hamiltonian yields a new pseudo-spectral basis $\left\{\ket{\widetilde{\psi\,}_a^\mathrm{L/R}}\right\}_{{n}_\mathrm{L}\equiv {n}_\mathrm{R}}$, where the Hamiltonian takes the form
\begin{equation}\begin{aligned}\label{eq:H_dimer}
  \widetilde{\matr{h}}_\mathrm{dimer}
    &= \begin{pmatrix}
        \widetilde{\matr{H}}_\mathrm{L} & \widetilde{\matr{T}}  \\
        \widetilde{\matr{T}}^\dagger & \widetilde{\matr{H}}_\mathrm{R}
       \end{pmatrix}.
\end{aligned}\end{equation}
After bringing in phase the left and right lead basis functions $\big(\left\{\ket{\widetilde{\psi\,}_a^\mathrm{L}}\right\}_{{n}_\mathrm{L}}$ and $\left\{\ket{\widetilde{\psi\,}_a^\mathrm{R}}\right\}_{{n}_\mathrm{R}}$, respectively$\big)$ 
the matrix elements  in Eq.~\eqref{eq:H_dimer} obey the symmetry relations
$\widetilde{\matr{H}}_\mathrm{L}\equiv \widetilde{\matr{H}}_\mathrm{R}$ and $\widetilde{\matr{T}}\equiv \widetilde{\matr{T}}^\dagger$.
That is, in energy space, the left and the right lead are equivalent. 
Further, the diagonal blocks $\widetilde{\matr{H}}_\mathrm{L/R}$ of the Hamiltonian are diagonal,
with their entries containing the associated eigenvalues.

Now, the eigenfunctions of the dimer can be transformed to the original system basis (cf. Eq.~\eqref{eq:H_KS})
using the resolution-of-identity
\begin{eqnarray}
  \ket{{\psi\,}_a^\mathrm{L/R}} &=& \sum_b^{{{n}_\mathrm{MO}}} \ket{\varphi_b^{\phantom{|}}} \matr{U}_\mathrm{L/R}^{(b,a)}\\
    \matr{U}_\mathrm{L/R}^{(b,a)} &=& \dotprod{\varphi_b}{\widetilde{\psi\,}_a^\mathrm{L/R}}\label{eq:basis_trafo}
\end{eqnarray} 
To ensure that subsequent quantum dynamics simulations remain computationally tractable,
we choose to retain only subsets of ${n}_\mathrm{L/R}$ lead states and ${n}_\mathrm{MO}$ molecular orbitals
within a symmetric energy window around the Fermi energy.
This gives rise to two convergence parameters: 
the energy range $\Delta E_\mathrm{lead}$ for choosing the lead basis functions, ${n}_\mathrm{L/R}$,
and the energy range $\Delta E_\mathrm{basis}$ for choosing the basis set size of the complete system, ${n}_\mathrm{MO}$.
The convergence of the dynamics with respect to these two parameters is benchmarked in the next section.
This procedure allows defining two transformation matrices for the left and right localized lead states,
$\matr{U}_\mathrm{L/R}$, with elements given by Eq.~\eqref{eq:basis_trafo}.

\subsubsection{Defining Localized States of the Extended Molecule}
As a final step towards the construction of the localized Hamiltonian,
the extended molecule (M) pseudo-spectral basis functions 
are localized according to the linear metric
\begin{eqnarray}\label{eq:loc_m}
  \big(\matr{1}-\matr{R}-\matr{L}\big)\matr{Q}_\mathrm{M} = \matr{Q}_\mathrm{M}\Lambda_\mathrm{M}
\end{eqnarray}
The projectors are defined as in Eq.~\eqref{eq:loc_lr_dimer}, but the matrix elements are computed using the delocalized MOs of the complete device model and not those of the dimer.
The ${n}_\mathrm{L/R}$ eigenfunctions associated with the smallest eigenvalues can be assigned to the left
and right leads. 
The remaining ${n}_\mathrm{M}=({n}_\mathrm{MO}-{n}_\mathrm{R}-{n}_\mathrm{L})$ largest eigenvalues
$\Lambda_\mathrm{M}$ are attributed to the extended molecule.
Diagonalization of the associated ${n}_\mathrm{M}\times{n}_\mathrm{M}$ Hamiltonian matrix 
allows defining a last transformation matrix as 
\begin{equation}\label{eq:extM}
{\matr{h}}_\mathrm{M}{\matr{U}}_\mathrm{M}={\matr{U}}_\mathrm{M}\matr{H}_\mathrm{M}\\
\end{equation}
where the matrix $\matr{H}_\mathrm{M}$ contains the eigenvalues of the extended molecule
pseudo-spectral basis functions.
The Hamiltonian of the finite nanojunction model in the localized pseudo spectral basis 
can be obtained by transforming the eigenvalue matrix of the extended system,
$\matr{\varepsilon}=\mathrm{diag}(\varepsilon_1,\varepsilon_2,\dots,\varepsilon_{\mathrm{MO}})$,
as follows
\begin{equation}\begin{aligned}\label{eq:H_sys}
  \matr{H}_\mathrm{sys} &= \matr{U}_\mathrm{sys}^\dagger\ \matr{\varepsilon}\ \matr{U}_\mathrm{sys}\\
    &= \begin{pmatrix}
        {\matr{H}}_\mathrm{L} & {\matr{V}}_\mathrm{LM} & 0 \\
        {\matr{V}}_\mathrm{LM}^\dagger & \matr{H}_\mathrm{M}& {\matr{V}}_\mathrm{MR} \\
        0 & {\matr{V}}_\mathrm{MR}^\dagger & {\matr{H}}_\mathrm{R}
       \end{pmatrix}
\end{aligned}\end{equation}
The rectangular matrices $\matr{V}_\mathrm{LM}$ and $\matr{V}_\mathrm{MR}$ describe the couplings from the extended molecule to the respective lead, in the local pseudo-spectral basis.
The total unitary transformation, $\matr{U}_\mathrm{sys}$, 
allows to numerically define an orthonormal set of pseudo-spectral one-electron basis functions from a subset of MOs obtained from standard quantum chemistry calculations.
It takes a block diagonal form
\begin{equation}\begin{aligned}
  \matr{U}_\mathrm{sys}=
       \begin{pmatrix}
       \matr{U}_\mathrm{L} & 0 & 0 \\
       0 & \matr{U}_\mathrm{M}& 0 \\
       0 & 0 & \matr{U}_\mathrm{R} 
       \end{pmatrix}
\end{aligned}\end{equation}
This yields three subsets of basis functions for 
the left lead $\left\{\ket{\psi_a^\mathrm{L}}\right\}_{n_\mathrm{L}}$, the right lead $\left\{\ket{\psi_a^\mathrm{R}}\right\}_{n_\mathrm{R}}$, and the extended molecule $\left\{\ket{\psi_a^\mathrm{M}}\right\}_{n_\mathrm{M}}$, which are then used in dynamical simulations and their analysis.

\subsubsection{Parametrization of a Tight-Binding Model}
Since the number of transport channels is limited by the number of available lead states, the resolution of the electronic current at a given bias voltage can be quite low.
To circumvent this issue, we extend the leads by parameterizing a tight-binding Hamiltonian,
$\matr{H}_\mathrm{tb}$ from the elements of the lead dimer, Eq.~\eqref{eq:H_dimer}.
The procedure is sketched in the bottom panel of Fig.~\ref{fig:concept}.
First, the lead diagonal blocks of the Hamiltonian \eqref{eq:H_sys} are replaced by the dimer
diagonal blocks, ${\matr{H}}_\mathrm{L/R} \approx \tilde{\matr{H}}_\mathrm{L/R}$.
The off-diagonal blocks $\tilde{\matr{T}}$ describing the coupling between two lead units in energy space
are then used to add a new unit to the lead. This procedure can be repeated until convergence of the current
through the nanojunction is obtained.

The new lead blocks are further divided into two groups:
those belonging to a buffer region, ${\matr{H}}_\mathrm{L/R}^\mathrm{buff}$,
and lead units ${\matr{H}}_\mathrm{L/R}^\mathrm{lead}$.
Only the latter are coupled to the implicit electronic reservoir in order to enhance the resolution of the electronic current.
The former are assigned to the extended molecule region and contribute significantly to the convergence of the DLvN towards NEGF reference calculations 
by avoiding direct coupling between the electronic reservoir and the scattering region.
Note that, for all non-equilibrium Green's function (NEGF) reference calculations shown in this paper,
the tight-binding Hamiltonian serves directly as input. This simplifies comparison of the currents obtained
via the DLvN and NEGF formalisms.

For the propagation using the DLvN formalism, $\matr{H}_\mathrm{tb}$ is brought into the form of $\matr{H}_\mathrm{sys}$ (cf. Eq.~\eqref{eq:H_sys}) by diagonalization of the extended molecule, the left and right lead blocks.
For clarity, we will refrain from introducing a new symbol for this final Hamiltonian here.
Instead, we will refer to Eq.~\eqref{eq:H_sys} hereafter.
Since all states are propagated explicitly in the DLvN equation, extension of the tight binding Hamiltonian greatly 
increases the associated computational effort. 
We observed that pruning the MO basis at this stage, as proposed in elsewhere~\cite{ZelovichHod2016nonorthogonal},
reduces the numerical effort at the expense of a violation of the Pauli principle.

\subsection{Monitoring the Electron Dynamics}

In the DLvN formalism, the time-evolution of the  block of the density matrix corresponding to the extended molecule
can be written by exploiting the structure of the localized Hamiltonian Eq.~\eqref{eq:H_sys} as follows
\begin{equation}\begin{aligned}\label{eq:DLvN_M}
  \frac{\partial \matr{\rho}_\mathrm{M}(t)}{\partial t} = &-\frac{\imath}{\hbar}\left[\matr{H}_\mathrm{M},\matr{\rho}_\mathrm{M}(t)\right] \\
    &-\frac{\imath}{\hbar}\left(\matr{V}_\mathrm{ML}\matr{\rho}_\mathrm{LM}(t)-\matr{\rho}_\mathrm{ML}(t)\matr{V}_\mathrm{LM}\right)  \\
    &-\frac{\imath}{\hbar}\left(\matr{V}_\mathrm{MR}\matr{\rho}_\mathrm{RM}(t)-\matr{\rho}_\mathrm{MR}(t)\matr{V}_\mathrm{RM}\right),
\end{aligned}\end{equation}
Taking the trace, $\tr{}{\frac{\partial \matr{\rho}_\mathrm{M}(t)}{\partial t}}$, yields the temporal change of
the total number of electrons in the extended molecule.\cite{ZelovichHod2016nonorthogonal}
This quantity comprises three contributions:
i) the probability flux within the extended molecule, 
ii) the probability flux from the left lead to the central unit (the influx),
and iii) the probability flux from the right lead to the central unit (the outflux).
Under steady-state conditions, the probability flux within the extended molecule vanish,
as well as the sum of the latter two contributions.
Thus, the total number of electrons in the extended molecule stays unchanged.
Multiplying the influx and the outflux by the elementary charge $e$, yields the current per spin channel
\begin{equation}\begin{aligned}
  I_\mathrm{LM}(t) &= &\frac{2e}{\hbar}&\sum_a^{N_\mathrm{L}}\sum_b^{N_\mathrm{M}} \matr{V}_\mathrm{LM}^{(a,b)}\,\im{\matr{\rho}_\mathrm{LM}^{(a,b)}(t)} \\
  I_\mathrm{MR}(t) &= -&\frac{2e}{\hbar}&\sum_a^{N_\mathrm{R}}\sum_b^{N_\mathrm{M}} \matr{V}_\mathrm{RM}^{(a,b)}\,\im{\matr{\rho}_\mathrm{RM}^{(a,b)}(t)}.
\end{aligned}\end{equation}
The average of these values describes the net current passing through the junction
\begin{equation}\begin{aligned}\label{eq:I_t}
  I(t) = \left(I_\mathrm{LM}(t) + I_\mathrm{MR}(t)\right)/2,
\end{aligned}\end{equation}
where for a steady state, the condition $I_\mathrm{LM}(t)=I_\mathrm{MR}(t)$ holds.

Alternatively, the local current in the scattering region can be extracted from the time-dependent density operator,
which in the basis of the localized eigenstates takes the following form
\begin{equation}\begin{aligned}
\hat{\matr{\rho}}(t)=\sum_{a,b}^{n_\mathrm{MO}} \matr{\rho}^{(a,b)}(t) \ket{\varphi_a}\bra{\varphi_b}.
\end{aligned}\end{equation}
To compute the local current in the region of space localized on the extended molecule,
it suffice to project the driven Liouville von-Neumann equation Eq.~\eqref{eq:DLvN} in
position representation
\begin{equation}\begin{aligned}\label{eq:DLvN_pos}
  &\frac{\partial\expect{\vec{r}\,}{\hat{\matr{\rho}}(t)}{\vec{r}\,}}{\partial t} 
  \\&\quad= 
  -\frac{\imath}{\hbar}\expect{\vec{r}\,}{\left[\hat{H},\hat{\matr{\rho}}(t)\right] }{\vec{r}\,}
  -\frac{\imath}{\hbar}\expect{\vec{r}\,}{\left[\imath \hat{W},\hat{\matr{\rho}}(t)\right]_+ }{\vec{r}\,}\\
   &\quad= -\frac{\imath}{\hbar}\expect{\vec{r}\,}{\left[-\frac{\hbar^2}{2m_e}\nabla_e^2+\hat{v}_\mathrm{KS},\hat{\matr{\rho}}(t)\right] }{\vec{r}\,} + \mathcal{W}(\vec{r},t),\\
\end{aligned}\end{equation}
Since the effective potential $\hat{v}_\mathrm{KS}$ is a multiplicative operator in position representation and the complex Hamiltonian $\imath \hat{W}$ acts only onto the lead basis functions, Eq.~\eqref{eq:DLvN_pos} simplifies to the electronic continuity equation for the extended molecule volume
\begin{equation}\begin{aligned}\label{eq:ceq}
  \frac{\partial \rho_\mathrm{M}(\vec{r},t)}{\partial t} &= -\vec{\nabla}_e\cdot \vec{j\,}_\mathrm{M}(\vec{r},t),
\end{aligned}\end{equation}
where the time derivative of the electron density is referred to as electronic flow.
Note that, since the lead eigenfunctions,
$\left\{\psi_a^\mathrm{L/R}\big(\vec{r}\,\big)\right\}_{n_\mathrm{L/R}}$
are negligibly small in the scattering region due to the localization procedure,
the basis functions introduced by the tight-binding extension can also be safely ignored.
Hence, the time-dependent electronic (probability) flux density at a given point localized
in the scattering region is given by
\begin{equation}\begin{aligned}\label{eq:je}
  \vec{j\,}_\mathrm{M}(\vec{r},t) &=  \sum_{a<b} 2\imath\,\im{\matr{\rho}^{(a,b)}_\mathrm{M}}\vec{j\,}_\mathrm{M}^{(a,b)}(\vec{r}\,)\\
\end{aligned}\end{equation}
where the time-independent state-to-state electronic flux density is defined as
\begin{equation}\begin{aligned}
  &\vec{j\,}_\mathrm{M}^{(a,b)}(\vec{r},t) 
  \\&\quad=- \frac{\imath\ \hbar}{2m_e} \left(\varphi_a^\mathrm{M}(\vec{r}\,) \vec{\nabla}_e \varphi_b^\mathrm{M}(\vec{r}\,) - \varphi_b^\mathrm{M}(\vec{r}\,) \vec{\nabla}_e \varphi_a^\mathrm{M}(\vec{r}\,)\right). 
\end{aligned}\end{equation}
For consistency with the current definition $I(t)$, Eq.~\eqref{eq:ceq} is multiplied with the elementary charge
to obtain the continuity equation for charge conservation relating the electronic charge density, $e\cdot\rho_\mathrm{M}(\vec{r},t)$, with the electronic current density, $\vec{J\,}_\mathrm{M}(\vec{r},t)=e\cdot\vec{j\,}_\mathrm{M}(\vec{r},t)$.
Interestingly, the negative integral over $\vec{J\,}_\mathrm{M}(\vec{r},t)$ corresponds to the electronic dipole moment in velocity gauge.\cite{nafie1997electron,detCIorbkit_I}

When representing the above quantities in position representation, 
they can be used to estimate the convergence of the electronic continuity in position space.
Those relations are derived and benchmarked in Sec.~B of the Supporting Information. 

\section{Computational Details}

All quantum chemical calculations were performed using \hbox{\textsc{TURBOMOLE}} \cite{TURBOMOLE} 
at the density functional theory (DFT) level of theory using the PBE0\cite{Adamo1999pbe0}
hybrid functional and a def2-SVP basis set\cite{def2}.
Three structures have been considered: the ON conformer ($\theta\approx 0^{\circ}$, cf. Fig.~\ref{fig:switch_sketch}) 
and the OFF conformer ($\theta\approx 90^{\circ}$, cf. Fig.~\ref{fig:switch_sketch}) of the molecular junction as depicted in Fig.~\ref{fig:concept} (upper panel),
as well as the lead dimer, as depicted in Fig.~\ref{fig:concept} (central panel).
The size of the lead sections in the extended molecule section is chosen large enough such that the influence of the central nitrophenyl group on the leads is negligible.
Further, the mechanistic details of the spatially resolved current dynamics can be investigated on a larger part of the so-called ``extended'' molecule.
Note that, in previous work, the NEGF reference was shown to be already converged with smaller leads.\cite{AgapitoCheng2007florida,PohlTremblay2016opegnr}

The reference current-voltage characteristics ($I-V$ curve) of the molecular switching device was obtained from the tight-binding Hamiltonian depicted in Fig.~\ref{fig:concept} (lower panel)
using the non-equilibrium Green's function  method and the Landauer-B\"uttiker formalism,
as implemented in \textsc{ASE}\cite{ASE,Larsen2017ASE,Thygesen2003ASEtransport,Thygesen2005ASEtransport,Strange2008ASEtransport}.
This implementation has also been used to obtain the lead's self-energy (cf. Eq.~\eqref{eq:broadening}).
The propagation of the reduced density matrix was performed at $T=0\,\mathrm{K}$ using \textsc{gloct}\cite{jcTS2008rho_TDCI},
an in-house implementation of a Markovian master equation propagator based on a preconditioned adaptive step size
Runge-Kutta algorithm \cite{04:TC:passrk}. 
In the simulation, the time step size was found to vary between $\Delta t = 0.001\,\mathrm{as}$ and $\Delta t = 150\,\mathrm{as}$ with an average value of $\Delta t = 30\,\mathrm{as}$.

For the computation of the electron density and electronic current density, the molecular orbitals
and the spatial derivatives thereof were first projected on a grid using \textsc{ORBKIT}\cite{orbkit}.
These quantities were combined with the time-dependent coefficients of the reduced density matrix
with our open-source \textsc{Python} package \textsc{detCI@ORBKIT}\cite{detCIorbkit_I,detCIorbkit_II}.
The results were visualized using \textsc{Matplotlib}\cite{Hunter2007Matplotlib}.
All streamline plots of the current density were created using \textsc{Amira}\cite{Amira}.
Here, a few hundred streamlines are seeded in the volume surrounding the left lead region (blue box in Fig.~\ref{fig:concept} (upper panel)) for positive 
and the right lead region (orange box in Fig.~\ref{fig:concept} (upper panel)) for negative bias voltages according to the magnitude of the electron density in that volume.
The color and opacity of the streamlines is chosen according the magnitude of the current density. 
The depictions of the molecular structures in Fig.~\ref{fig:switch_sketch} and Fig.~\ref{fig:concept}
were created using \textsc{XCrySDen}\cite{xcrysden}.

\section{Results and Discussion}

\subsection{Convergence Behavior}
\subsubsection{Localization Procedure}
The localization procedure depends on two energy parameters:
$\Delta E_\mathrm{lead}$ for choosing the lead basis functions  and  $\Delta E_\mathrm{basis}$ for choosing the basis set of the complete system.
\begin{figure}[t!]
\centering\includegraphics[width=0.9\linewidth]{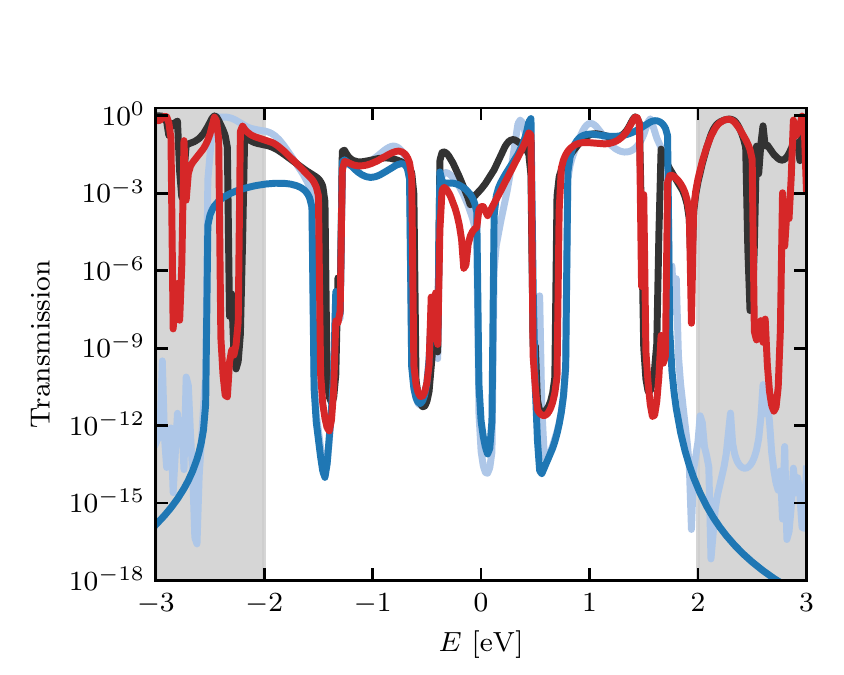}
\caption{\label{fig:T_E}
The transmission function within the energy range of interest (conductance around the Fermi energy
within an energy window of  $U_\mathrm{max}=4\,\mathrm{V}$) for different choices of the two energy parameters used within the localization procedure:
$\Delta E_\mathrm{lead}$ for choosing the lead basis functions  and  $\Delta E_\mathrm{basis}$ for choosing the basis set of the complete system.
Dark blue line: the minimum energy ranges ($\Delta E_\mathrm{lead}=4\,\mathrm{eV}$, $\Delta E_\mathrm{basis}=4\,\mathrm{eV}$),
light blue line: the minimal choice for the size of the lead basis ($\Delta E_\mathrm{lead}=4\,\mathrm{eV}$, $\Delta E_\mathrm{basis}=80\,\mathrm{eV}$), 
red line: the energy range used in this work ($\Delta E_\mathrm{lead}=6\,\mathrm{eV}$, $\Delta E_\mathrm{basis}=8\,\mathrm{eV}$),  
and black line: a reference ($\Delta E_\mathrm{lead}=40\,\mathrm{eV}$, $\Delta E_\mathrm{basis}=80\,\mathrm{eV}$).
}
\end{figure}
To estimate the convergence of the localization procedure,
Fig.~\ref{fig:T_E} reports the influence of the energy windows on the transmission function
within the energy range of interest.
For the system investigated, we focus on the conductance around the Fermi energy
within an energy window of  $U_\mathrm{max}=4\,\mathrm{V}$.
The prominent feature around the Fermi level observed in the reference NEGF results can only be reproduced
accurately using a very large basis ($\Delta E_\mathrm{lead}=40\,\mathrm{eV},~\Delta E_\mathrm{basis}=80\,\mathrm{eV}$, black line).
For the smallest possible energy windows ($\Delta E_\mathrm{lead}=4\,\mathrm{eV},~\Delta E_\mathrm{basis}=4\,\mathrm{eV}$, dark blue line) qualitatively meaningful results are only obtained between $-2\,\mathrm{eV}$ and $2\,\mathrm{eV}$.
Using a very large window for the resolution-of-identity, $\Delta E_\mathrm{basis}=80\,\mathrm{eV}$, while 
keeping the number of lead states small ($\Delta E_\mathrm{lead}=4\,\mathrm{eV}$, light blue line).
does not improve the appearance of the conductance curve.
On the contrary, a moderate increase of the lead energy window ($\Delta E_\mathrm{lead}=6\,\mathrm{eV}$)
and of the resolution-of-identity ($\Delta E_\mathrm{basis}=8\,\mathrm{eV}$)
allows to recover all features of the reference (see red line in Fig.~\ref{fig:T_E}) at a tractable numerical cost.
These are the parameters used throughout this work, which gives rise to $n_\textrm{M}=52$ localized
basis function in the scattering region and $n_\textrm{L/R}=9$ lead functions, without considering any tight-binding extension. 

\begin{figure*}[t]
\centering\includegraphics[width=0.9\linewidth]{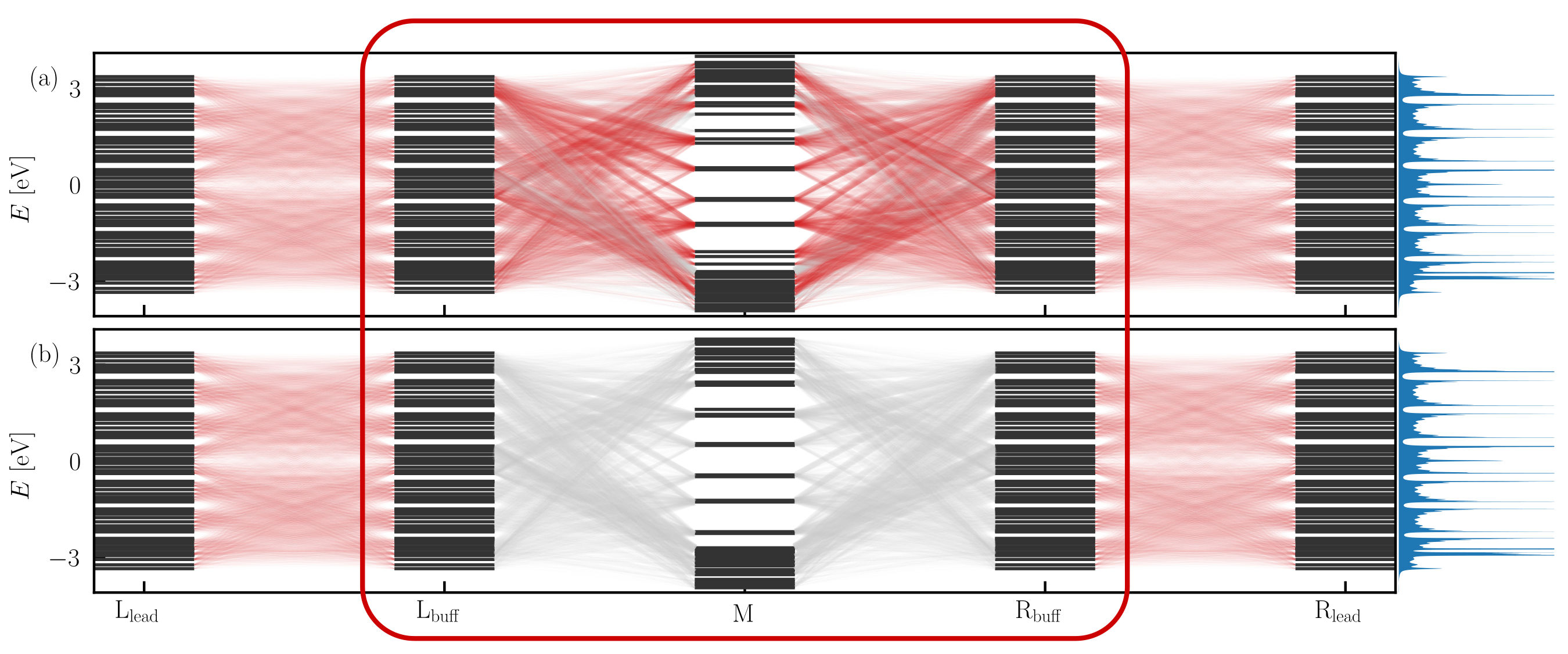}
\caption{\label{fig:energy_levels}
Energy levels (black horizontal lines) and their couplings (red and grey lines) for the ON (upper panel) and the OFF conformer (lower panel) extended with ten buffer and ten lead units
showing the three parts of the junction: 
the left lead ($\mathrm{L}_\mathrm{lead}$), the extended molecule ($\mathrm{L}_\mathrm{buff}+\mathrm{M}+\mathrm{R}_\mathrm{buff}$, red box), 
and the right lead ($\mathrm{R}_\mathrm{lead}$).
The extended molecule (red box) consists of left and right buffer units $\mathrm{L}_\mathrm{buff}$ and $\mathrm{R}_\mathrm{buff}$ and 
the original extended molecule region (M, cf. black solid box in Fig. \ref{fig:concept} (upper panel)).
It is treated as the coherent scattering region in Eq.~\eqref{eq:H_sys}.
The linewidth of the connectors between states is chosen according to the coupling strength.
To allow identifying the conducting states, the connectors are colored in red if the coupling of a specific state of the extended molecule 
with both, the left and the right buffer units, differs by at most a factor of 2 for the largest component on each lead.
The density of states of the leads, depicted as blue shaded curve at the abscissa, is broadened with a Lorentzian
of the same width as in the subsequent DLvN propagations (cf. Eq.~\eqref{eq:broadening}).
The depiction of the diagonal elements of the Hamiltonian blocks as energy levels and their couplings as lines connecting these levels to visualize the structure of the Hamiltonian was already proposed in Ref.~\cite{ZelovichHod2015real_vs_state_space,ZelovichHod2016nonorthogonal}.
}
\end{figure*}

\subsubsection{Tight-Binding Model}
Using the coupling elements of the Hamiltonian matrix of the lead dimer $\tilde{\matr{T}}$,
the system Hamiltonian can be extended at will 
by adding additional buffer $\tilde{\matr{H}}_\mathrm{L/R}^\mathrm{buff}$
and lead units $\tilde{\matr{H}}_\mathrm{L/R}^\mathrm{lead}$.
Fig.~\ref{fig:energy_levels} shows the spectrum of the resulting Hamiltonian for the ON (upper panel) and
OFF (lower panel) conformations for an exemplary system with ten buffer and ten lead units. 
The black horizontal lines refer to the energy levels of the different regions of the nanojunction, 
and the connectors between these energy levels correspond to the couplings between the pseudo-spectral states of the different regions, e.g., $\matr{V}_\mathrm{L_{buff}M/R_{buff}M}$.
It can be observed that the energy spectrum of both logical states is nearly identical.
While the spectrum of the extended molecule is dense at low and at high energies, the pseudo-eigenstates of both leads are more evenly distributed and form bands at intermediate energies,
with an energy spacing between the bands of $\Delta \varepsilon \approx 0.2\,\mathrm{eV}$.
At the ordinate, the  density of states (DOS) of the leads is plotted using the same Lorentzian broadening as defined by Eq.~\eqref{eq:broadening}.

The linewidth for interstate couplings in Fig.~\ref{fig:energy_levels} is chosen according to the strength of the respective coupling.
Interestingly, when we compare the ON and OFF conformations, not only the preferred coupling channels but also the coupling strengths are very similar.
To distinguish between conductive and non-conductive channels through the bridge,
we introduce a measure of the connectivity asymmetry of a particular molecular channels.
That is, the connectors are only colored in red if the coupling to a specific molecular channel differs by a factor of 2 at most for the largest coupling on each lead.
This reveals that the majority of the extended molecule states of the ON conformation are conductive,
while for the OFF conformation, nearly all states are asymmetrically coupled to the leads and therefore non-conductive.
The explanation can be found by analyzing the coupling to the (nearly) degenerate pairs of extended molecule states in more detail.
For each pair of states localized on the extended molecule, the coupling strength is approximately
the same with both leads in the ON conformation.
For the OFF logical state, one state of the doublet couples exclusively to the left while the other couples exclusively to the right lead.
Thus, it can be anticipated that the OFF conformation will be significantly less conducting than the ON logical state, even without performing any dynamical simulation.

\subsection{Modeling the Time-Dependent Electronic Current}

In the following, the DLvN approach is applied to investigate the electronic current in the OPE-GNR nanojunction model.
A linear voltage ramp from $U=0\,\mathrm{V}$ ($\mu_\mathrm{L/R}=0$) at $t=0$ to $U=4\,\mathrm{V}$ ($\mu_\mathrm{L/R}=\pm 2\,\mathrm{eV}$) at $t=4\,\mathrm{ps}$ is chosen to drive the dynamics.
At the beginning of the simulations, all subsections of the molecule are in thermal equilibrium {\it locally}.
Since this is not the thermal equilibrium of the total system, an ultrafast equilibration dynamics occurs in the early stages
of the simulation, as the coupling between the different parts of the nanojunction is suddenly switched on.
To avoid artifacts coming from this unphysical behavior, the system is first left to equilibrate for $1\,\mathrm{ps}$
before the bias voltage is ramped up slowly from the new initial time $t=0$.
Note that the Pauli principle is satisfied throughout the entire simulation for all setups investigated (cf. Sec.~A of the Supporting Information for details)

\begin{figure}[t!]
\centering\includegraphics[width=0.85\linewidth]{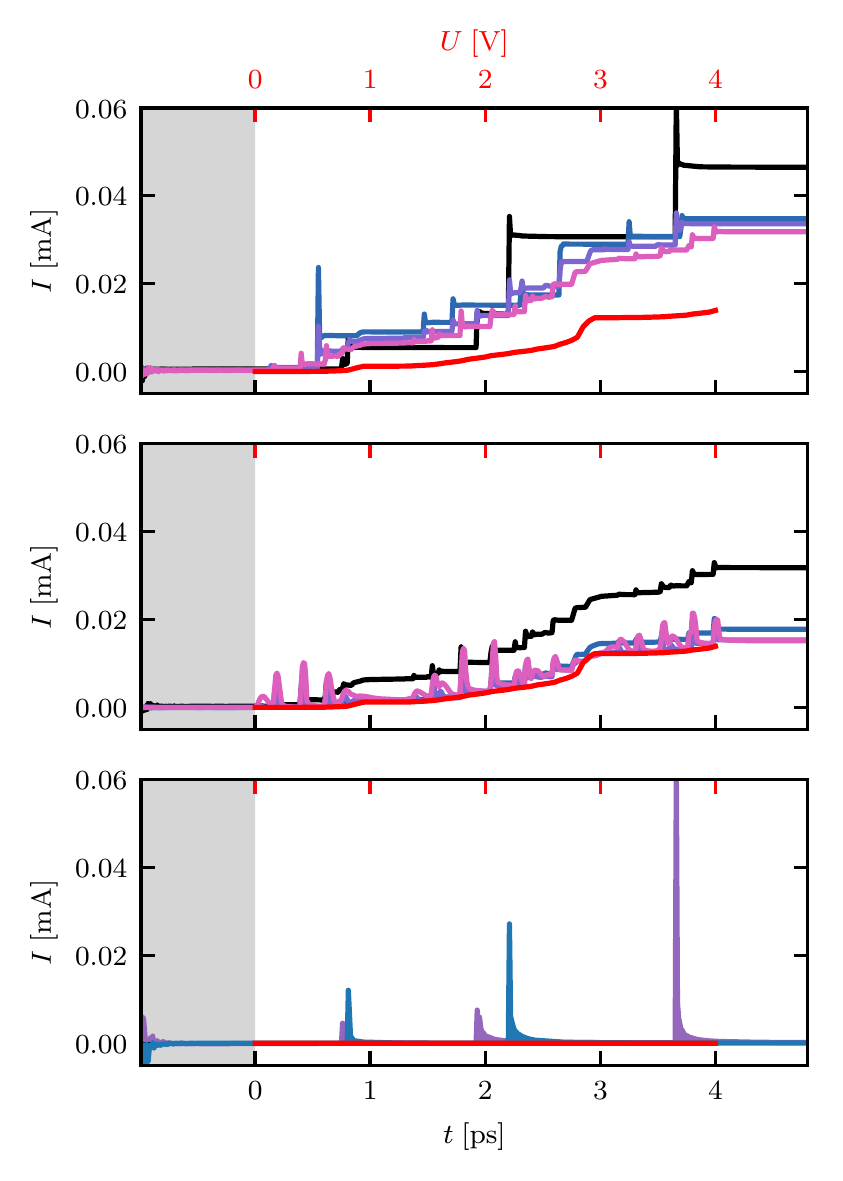}
\caption{\label{fig:iv_of_t}
Time-dependent current-voltage characteristic ($I-V$ curve) of the OPE-GNR junction $I(t)$ applying a linear voltage ramp from $U(t=0\,\mathrm{ps})=0$ to $U(t=4\,\mathrm{ps})=4\,\mathrm{V}$ for ON (upper and central panel),
and for OFF (lower panel) compared with NEGF reference calculations (red lines).
The grey shaded area highlights the equilibration time without bias voltage to account for the unphysical behavior at the beginning of the dynamics. 
Upper panel: The time-dependent current for the original system (cf. Fig. \ref{fig:concept} upper panel), and for two, five, and ten lead units  (color key: black, dark blue, purple, and pink, respectively).
Central panel: The time-dependent current for ten lead units with zero (black), one (dark blue), two (purple) and ten (pink) buffer units.
Lower panel: The two components contributing to the net current $I(t)$, i.e., the influx from the left lead ($I_\mathrm{LM}(t)$) and the negative outflux to the right lead ($-I_\mathrm{MR}(t)$), are plotted as blue and purple curves for the OFF conformer without tight-binding extension.
}
\end{figure}

Fig.~\ref{fig:iv_of_t} shows the time-evolution of the electronic current for both conformers at $T=0\,\mathrm{K}$
compared with the current voltage characteristics obtained from NEGF calculations (red curves).
Regarding the NEGF reference, it can be noticed that the current gradually increases
in smooth steps for the ON conformation (top and central panel of Fig.~\ref{fig:iv_of_t}), while the current for the OFF logical state (bottom panel of Fig.~\ref{fig:iv_of_t}) always remains very small.
At all potential biases, the ON/OFF current ratio lies between $10^2$ and $10^3$, which is in good agreement with
our previous findings\cite{PohlTremblay2016opegnr}.
The finite number of states within the time-dependent DLvN simulation restricts the effective applicable bias voltages to the available lead state energies.
For example, only four steps can be observed within the time-dependent current dynamics for the ON conformation (black curve in the upper panel of Fig.~\ref{fig:iv_of_t}).
The NEGF reference and the time-dependent results share the same qualitative features despite some marked deviations.
That is, both curves describe the same few transport channels showing up at the same bias voltages.

Whenever the time-dependent bias voltage hits a resonance in a lead state, a rapid rise in conductivity is observed followed by an equilibration to a lower lying plateau.
This rapid rise is observed for both the ON and OFF logical states.
To understand this phenomenon, the two contributions to the current, $I(t)$, (cf. Eq.~\eqref{eq:I_t}) are plotted separately
in the lower panel of Fig.~\ref{fig:iv_of_t} for the OFF configuration: 
the influx from the left lead $(I_\mathrm{LM}(t))$ as a blue and the negative outflux to the right lead ($-I_\mathrm{MR}(t)$) as a purple curve.
As can be seen from the figure, there is either an influx or an outflux in this OFF configuration, but never both simultaneously.
These dynamical features, which take place in the femtosecond time regime, can be associated with the population and
depopulation of extended molecular states reacting to the new boundary conditions.
Thus, those peak currents do not contribute to the overall current passing {\it through} this junction, and should 
be understood as an ultrafast equilibration response.
This ultrafast phenomenon will be further investigated in the next chapter from the perspective of the local current.

In order to enable a more precise description of the electric current dynamics for the ON state, 
both leads were extended as explained above (see Eq.~\ref{eq:basis_trafo}) by a certain number of tight binding units as buffer units between the central molecule and the leads and as additional lead units being coupled to the implicit electronic reservoir.
While the latter
allow for a higher resolution in the bias voltages and a better representation of the density of states in the leads,
the former prevents the direct coupling between the central unit and the implicit electron reservoir.\cite{ZelovichHod2015real_vs_state_space}
For the ON conformation, the results for two (dark blue), five (purple), and ten (pink) lead units are shown in the upper panel of Fig.~\ref{fig:iv_of_t}.
It can be recognized that, with increasing number of lead units, the large jumps in $I(t)$ between the different plateaus are gradually replaced by smoother transitions, and the curves converge slowly to reproduce the shape of the NEGF reference.
Moreover, the size of the peak currents due to the ultrafast equilibration dynamics is significantly reduced due to the
smaller energy gap between the states.
Interestingly, simulations at higher temperatures without tight-binding extension yield similar $I-V$ current profiles. 
This is due to the smoother change in population as temperature increases, see Eq.~\eqref{eq:fermidirac}.

The central panel of Fig.~\ref{fig:iv_of_t} shows the influence of introducing buffer units, i.e., lead units that are not coupled to the electronic reservoir, 
at the example of ten lead units with zero (black), one (dark blue), two (purple) and ten (pink) buffer units.
As can be seen, introducing just a single buffer unit (dark blue curve) significantly improves the result, 
and by adding two or more buffer units the $I-V$ curve is basically converged to the NEGF reference (red curve). 
Interestingly, the more buffer units we introduce, the more pronounced are the features occurring when a new transport channel is opened.
This phenomenon can be explained by the simple fact that 
increasing the number of states in the buffer implies more phases needing to equilibrate.
This is a signature of the non-Markovian equilibration dynamics, an important feature of the DLvN formalism.

\subsection{Spatially-Resolved Current Dynamics}

\subsubsection{Constant Bias and the Onset of Current Dynamics}
\begin{figure*}[t!]
\centering\includegraphics[width=0.9\linewidth]{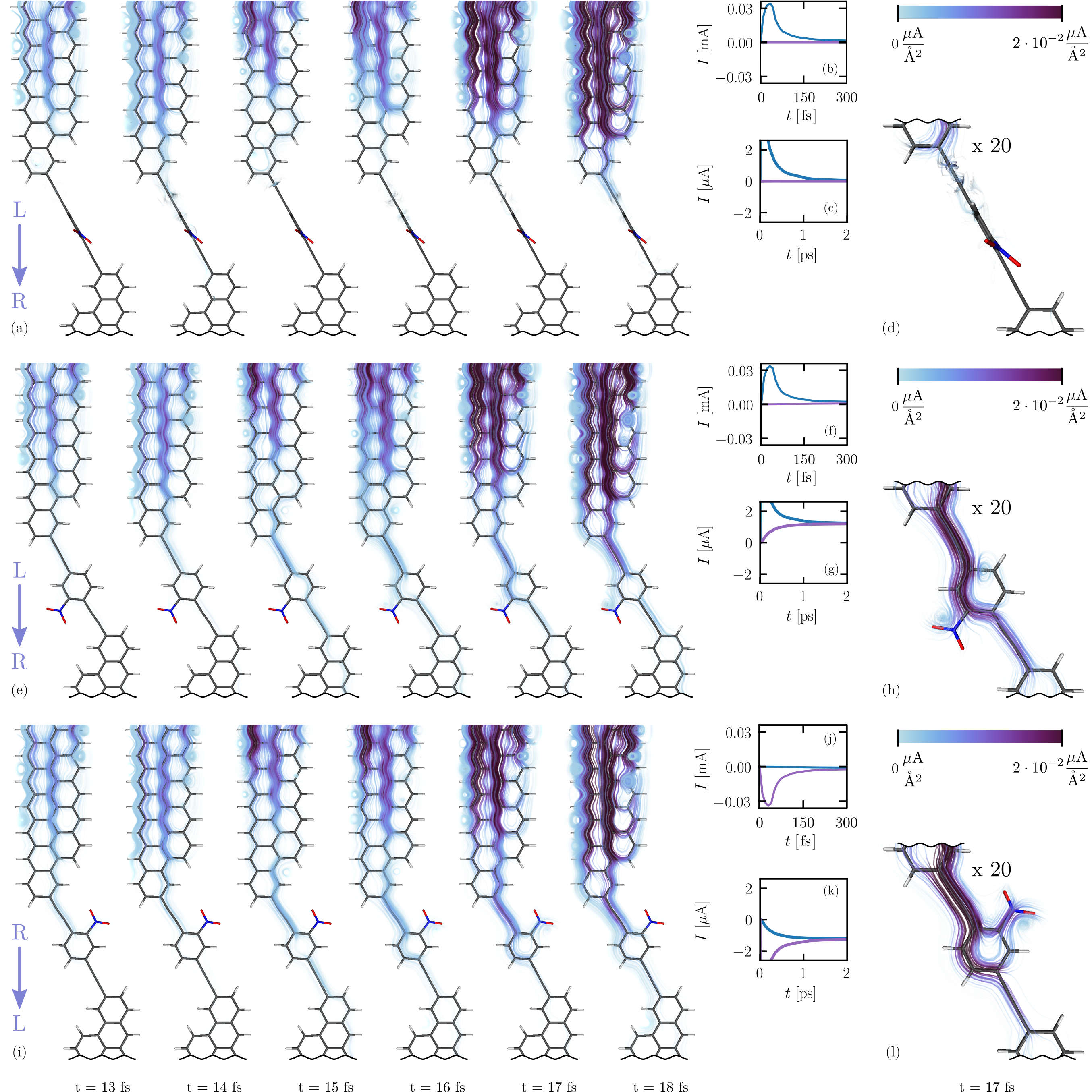}
\caption{\label{fig:je_05Va}
Streamline plots of the electronic current density on the extended molecule $J_\mathrm{M}(\vec{r},t)$   (in units of ${\rm\mu A}/\mathrm{\mathring{A}^2}$) for representative snapshots of the dynamics 
in the first $20\,\mathrm{fs}$ (a--d) for the OFF and (e--h) the ON conformations
with current coming from the left lead with $\mu_\mathrm{L}=+ 0.5\,\mathrm{eV}$, 
and additionally (i--l) for the ON conformation, a dynamics, where the current comes from the right lead with $\mu_\mathrm{R}=+ 0.5\,\mathrm{eV}$.
Before the potential bias was suddenly turned on at $t=0\,\mathrm{fs}$, the system was equilibrated for $1\,\mathrm{ps}$.
The streamlines are color-coded and their opacity is chosen according to the magnitude of $J_\mathrm{M}(\vec{r},t)$.
(d,h,l) Enlarged views for OFF and for the \textit{meta} and the \textit{ortho} scenario at $t=17\,\mathrm{fs}$, respectively.
(b,c), (f,g), and (j,k) Current--Voltage characteristics (I--V curves) for the different setups.  
The influx from the left lead $(I_\mathrm{LM}(t))$ is depicted as a blue and the outflux to the right lead $(I_\mathrm{MR}(t))$ as a purple curve.}
\end{figure*}

An important focus of this work is the investigation of the mechanistic details of the electron transport through the OPE-GNR nanojunction.
A natural choice for this analysis is the electronic current density, which provides a spatially resolved picture of the instantaneous flow of electrons.
Before regarding the electron dynamics of the linear voltage ramp, let us first consider the equilibration dynamics initiated 
when we suddenly switch on a bias voltage of $U=0.5\,\mathrm{V}$ on one side of the system.
Note that, as mentioned above, the system was first equilibrated for $1\,\mathrm{ps}$ at $U=0\,\mathrm{V}$
to avoid unphysical effects coming from thermalization.
Fig.~\ref{fig:je_05Va} shows the current density dynamics for this scenario for the extended Hamiltonian
using ten buffer and ten lead units.
It displays the first $20\,\mathrm{fs}$ for the OFF (cf. Fig.~\ref{fig:je_05Va}(a--d)) and the ON conformations
(cf. Fig.~\ref{fig:je_05Va}(e--h)) 
with current coming from the left lead with $\mu_\mathrm{L}=+ 0.5\,\mathrm{eV}$ ($\mu_\mathrm{R}= 0.0\,\mathrm{eV}$).
Additionally, for the ON conformation, a dynamics in which the current comes from the right lead with
$\mu_\mathrm{R}=+ 0.5\,\mathrm{eV}$ ($\mu_\mathrm{L}= 0.0\,\mathrm{eV}$, is depicted in
Fig.~\ref{fig:je_05Va}(i--l).
The current voltage statistics (cf. Figs.~\ref{fig:je_05Va}(b,c), \ref{fig:je_05Va}(f,g), and \ref{fig:je_05Va}(j,k)) show the same prominent features as discussed in the previous section: 
a rapid rise in current, following a slow exponential equilibration.
Recall that  the $I-V$ curves (cf. Eq.~\eqref{eq:I_t}) are calculated at the boundaries between the leads and the buffer units, i.e., ten buffer units away from the extended molecule. 
Further positive currents are defined as flowing from the left (L) to the right side (R) of the system.

An important feature of graphene nanojunctions such as the one studied here (cf. Fig.~\ref{fig:je_05Va}),
is that the current density originates from charge migration through $\pi$-molecular orbitals. 
This is consistent with our simulations, in which the electrons flow symmetrically above and
below the ZGNR plane and the current density in the molecular plane it is found to be strictly zero.
After switching on the bias voltage abruptly, it takes around $12\,\mathrm{fs}$ for the electrons
to reach the boundaries of the extended molecule, i.e. the scattering region.
This time delay correlates directly with the number of buffer units introduced in the tight-binding Hamiltonian.
The electronic current density propagates along the bonds preferably following the central pathway.
Within the next $3\,\mathrm{fs}$ it reaches the bridge connecting the ZGNR ribbon with the central molecule.
Since at that point, the current is constrained to flow along the ethynylene group, the electrons are scattered and partially reflected.
The resulting backflow interferes destructively with the inflow of electrons towards the central molecule, 
leading to a nearly vanishing current density in that region at $t\approx 15\,\mathrm{fs}$.
Interestingly, at twice this time ($t\approx 30\,\mathrm{fs}$), a drop in the I--V curve can be observed. 
This can be associated with reflected electrons arriving at the boundaries between the leads and the buffer unit,
where the current is calculated. 
Besides,
this backflow induces turbulences on the short edge (right edge) of the ZGNR ribbon.
This establishes a wide stationary eddy which becomes even larger as time increases.
While the current patterns on the incoming side of the lead are almost identical for ON and OFF within the first $20\,\mathrm{fs}$,
they differ significantly at the bridge and in the outgoing lead.
For the OFF conformation (cf. Fig.~\ref{fig:je_05Va}a), the route across the bridge is blocked because of the breakdown of the $\pi$-conjugation.
That is, a very small fraction of the current density  enters the central nitrophenyl group following a turbulent circular pathway (cf. zoomed view in Fig.~\ref{fig:je_05Va}d).
However, the vast majority is constantly reflected back to the incoming channel. 
In the later course of the dynamics (not shown), the shape of the current density patterns does not change significantly.
The eddy simply becomes slightly more pronounced at first, until the current density completely vanishes at $t\approx 500\,\mathrm{fs}$.
For the ON conformation (cf. Fig.~\ref{fig:je_05Va}(e,i)), a different picture emerges.
Although the electron flow is constrained by the bridge, it can reach the outgoing side already at $t=15\,\mathrm{fs}$ and continues propagating towards the outgoing lead.
This is facilitated by the delocalized $\pi$-system through the molecular bridge.

Let us now focus on the electron dynamics on the central nitrophenyl group, 
and investigate the influence of the nitro group on the charge migration mechanism.
While this group stands in \textit{meta} position relative to the incoming flux at positive biases,
$\mu_\mathrm{L}=+ 0.5\,\mathrm{eV}$ (cf. Fig.~\ref{fig:je_05Va}(e--h)), 
it is found in \textit{ortho} position for the reversed bias direction (cf. Fig.~\ref{fig:je_05Va}(i--l)).
In the first moments of the scattering event at $t=15\,\mathrm{fs}$,
the current densities seem to avoid the pathway along the nitro group for both polarities.
This changes drastically in the following few femtoseconds, when the electron withdrawing character of the nitro group becomes apparent.
Figs.~\ref{fig:je_05Va}h and \ref{fig:je_05Va}l show enlarged views of the \textit{meta} and \textit{ortho} scenarios for $t=17\,\mathrm{fs}$.
In the \textit{meta} case, the complete electronic current density is pulled towards the side bearing the nitro group.
This unilateral transport induces a backflow of the electronic current density along the opposite side of the central group.
Further, a fraction of the current density is pulled towards the inner oxygen of the nitro group,
where it establishes a small eddy.
In the \textit{ortho} case, a small share of the current density is dragged directly towards the nitro group, 
while the main part follows the opposite pathway along the central phenyl group and splits up again at the outlet of the central group.
From here, one part is flowing to the outgoing channel,
while the other part is flowing back towards the nitro group establishing an eddy at the inner oxygen -- in a similar manner as in the \textit{meta} case.
Recall that the nitro group is a \textit{meta}-directing group upon electrophilic aromatic substitutions.
It now stands in a \textit{meta} position with respect to the back-flowing electrons. 
As a consequence, a smaller amount of electronic current density is flowing towards the outgoing side of the molecule 
in the \textit{ortho} scenario (from the right to the left lead) than in the \textit{meta} case at that particular time.
Interestingly, this effect persists only for a few femtoseconds and does not alter the overall current voltage characteristics after equilibration.
This insightful result could be exploited to improve the nanojunction,
e.g., by replacing the nitro group  with an \textit{ortho}-directing halide.
We can hypothesize from the present simulations that this substitution could reduce the amount of current that flows 
back, thus reducing the turbulence through the device. It is understood that a more laminar flow of electrons
is a desirable quality of a nanojunction in its ON state, as it would improve its conductivity
and potentially reduce heat production.

\begin{figure*}[t!]
\centering\includegraphics[width=0.9\linewidth]{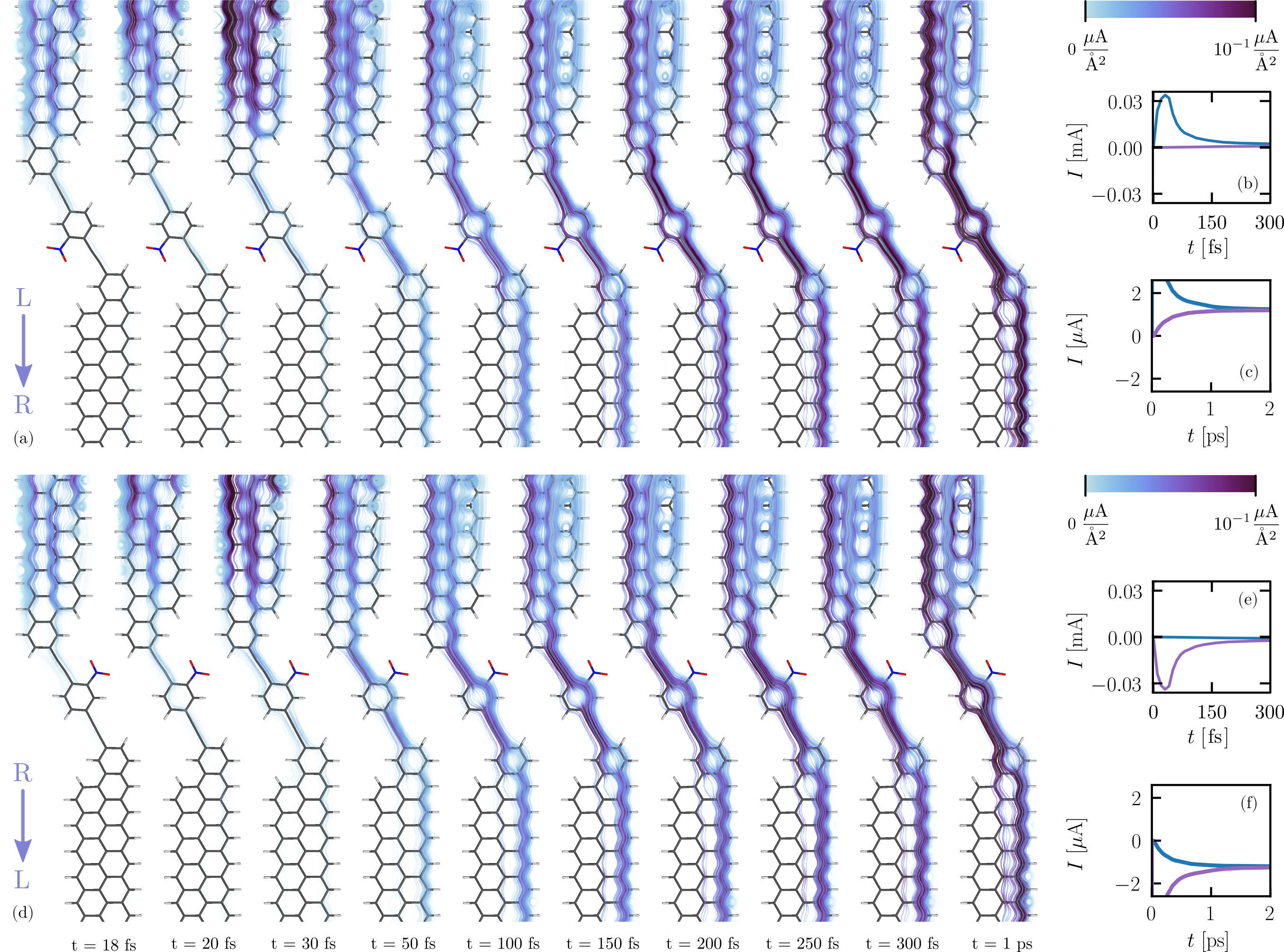}
\caption{\label{fig:je_05Vb}
Streamline plots of the electronic current density on the extended molecule $J_\mathrm{M}(\vec{r},t)$   (in units of ${\rm\mu A}/\mathrm{\mathring{A}^2}$) for representative snapshots of the dynamics 
(a--c) for the ON conformation
with current coming from the left lead with $\mu_\mathrm{L}=+ 0.5\,\mathrm{eV}$, 
and additionally (d--f) for the ON conformation, a dynamics, where the current comes from the right lead with $\mu_\mathrm{R}=+ 0.5\,\mathrm{eV}$.
Before the potential bias was suddenly turned on at $t=0\,\mathrm{fs}$, the system was equilibrated for $1\,\mathrm{ps}$.
The streamlines are color-coded and their opacity is chosen according to the magnitude of $J_\mathrm{M}(\vec{r},t)$.
The dynamics of the first $20\,\mathrm{fs}$ can be found in Fig.~\ref{fig:je_05Va}.
Please note the different colormaps.
(b,c) and (e,f) Current--Voltage characteristics (I--V curves) for the different setups.  
The influx from the left lead $(I_\mathrm{LM}(t))$ is depicted as a blue and the outflux to the right lead $(I_\mathrm{MR}(t))$ as a purple curve.}
\end{figure*}

Fig.~\ref{fig:je_05Vb} shows --- with a different scaling for the colormap --- the later equilibration dynamics for both scenarios starting from the last frame of Fig.~\ref{fig:je_05Va} at $t= 18\,\mathrm{fs}$.
On the incoming side of the ZGNR, the magnitude of the current density rises until
$t\approx 30\,\mathrm{fs}$ before it drops again. 
This feature is consistent with the maximum of the $I-V$ curve, and it
coincides with a change in the transport mechanism from a central pathway
towards an edge transport along the long side of the ribbon.
Further, it coincides with the establishment of a specific output channel, 
in which the outgoing current density flows laminarly along the long edge (right edge) of the ribbon.
Besides, the wide eddy established in the beginning of the dynamics on the incoming channel
gets more pronounced and persists even after equilibration.

Starting from $t> 50\,\mathrm{fs}$, the transport mechanism through the central group changes similarly for both 
current directions.
Now, the electron dynamics proceeds preferably along the right side of molecular junction,
independently of the position of the nitro group.
The \textit{meta}-directing influence of the nitro group on the transport mechanism can
nonetheless be observed. When the current flows from left to right (the \textit{meta} scenario),
a considerable fraction of the current still flows along the nitro group side of the ring.
This change in mechanism coincides with the timescale in which the magnitude of the current density
on the outgoing lead steadies and reaches a magnitude comparable to the incoming one.
The overall transport mechanism is almost converged at $t\approx 200\,\mathrm{fs}$.
Looking at the values of the incoming and the outgoing current at the extended molecule boundaries,
we can define a quasi-stationary condition when their magnitude differ by less than $10\,\%$.
As can be seen from the insets Fig.~\ref{fig:je_05Vb}c and Fig.~\ref{fig:je_05Vb}f,
this quasi-stationary condition is only reached at times $t\ge 1\,\mathrm{ps}$ after the potential
bias was switched on.
This is consistent with the picture emerging from the orbital populations, as shown in Sec.~A of the ESI$^\dag$.

\begin{figure*}[t!]
\centering\includegraphics[width=0.9\linewidth]{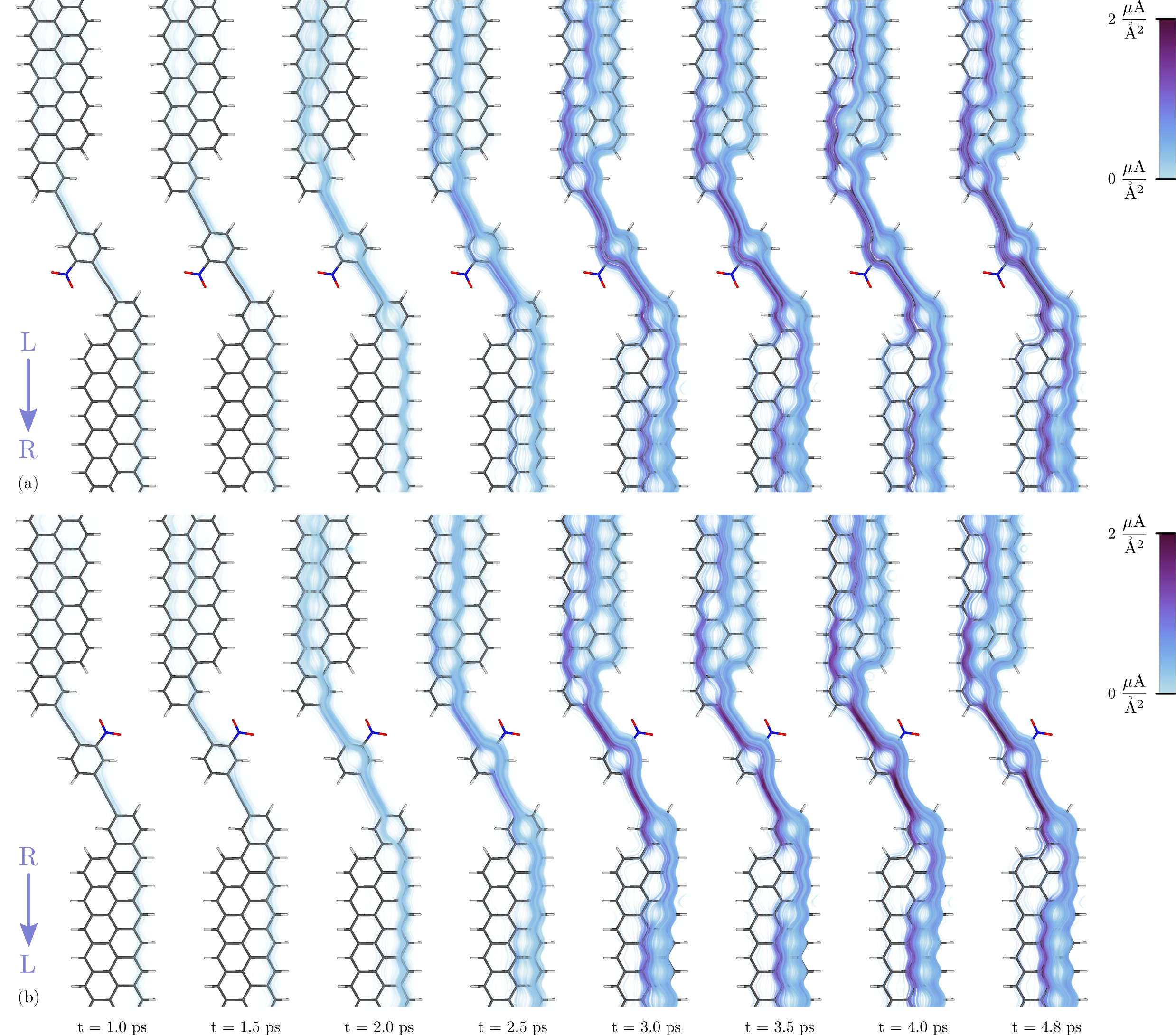}
\caption{\label{fig:je_4ps}
Streamline plots of the electronic current density on the extended molecule $J_\mathrm{M}(\vec{r},t)$   (in units of ${\rm\mu A}/\mathrm{\mathring{A}^2}$ for representative snapshots of (a) the dynamics 
shown in Fig.~\ref{fig:iv_of_t}a with a ten lead and ten buffer units tight-binding extension and (b) the same dynamics but with opposite sign.
The streamlines are color-coded and their opacity is chosen according to the magnitude of $J_\mathrm{M}(\vec{r},t)$.
Note that the current density plotted in this figure is a three-dimensional vector field and that smaller current density in the foreground cover up very large current density on the bonds.
This can lead to a situation, where some paths appear less favored than they really are, 
e.g., the current on the density on the bridge in Fig.~(b) at $t=4.8\,\mathrm{ps}$ is covered up by a very broad electron transport. 
}
\end{figure*}

\subsubsection{Time-Dependent Potential Bias}

Fig.~\ref{fig:je_4ps} (upper panel) shows representative snapshots of the electronic current density as streamline plots 
for a time-dependent potential bias.
A linear voltage ramp from $U=0\,\mathrm{V}$ ($\mu_\mathrm{L/R}=0$) at $t=0$ to $U=4\,\mathrm{V}$ ($\mu_\mathrm{L/R}=\pm 2\,\mathrm{eV}$) at $t=4\,\mathrm{ps}$ (cf. Fig.~\ref{fig:iv_of_t})
is applied to the ON conformation, described using a tight-binding Hamiltonian with ten buffer and ten lead units. 
The lower panel shows the current patterns for the same linear voltage ramp but in opposite direction (from R to L).
The first frame of the time series depicted in Fig.~\ref{fig:je_4ps} ($t=1\,\mathrm{ps}$) corresponds to the same chemical potential for the influx side of the system as in the previous example and, as such,
the mechanism at $1\,\mathrm{ps}$ is very similar.
That is, the current density flows laminarly along the long edges of the ZGNR ribbons and crosses the bridge on the right edge of the central group in a broad delocalized stream.
Moreover, a stationary eddy on the incoming side of the junction can be observed as well.
Its magnitude is too small to be seen with the colormap used in the figure.
As the bias voltage increases, the preferred path of the electron dynamics changes from an edge transport to a central pathway, 
first on the incoming side ($t\approx 2.5\,\mathrm{ps}$) and subsequently ($t\approx 3.0\,\mathrm{ps}$)
also on the outgoing side.
Moreover, the stationary eddy on the short edge of the incoming side vanishes and is replaced by a transport channel following the edge down to the central group.
In general for large bias voltages,
most of the current density propagates along the bonds on a meandering path from the incoming to the outgoing side of the device.
As can be expected from classical fluid dynamics, the current density is largest at the bottlenecks of the nanojunction,
i.e., at the triple bonds connecting the leads with the nitrophenyl group.
The current density remains large along the imaginary line that extends the axis spanned by these bonds until it reaches the edge of the nanoribbon. 
The transmission axis along the molecular junction does not align with the overall direction of the electron transport in the nanojunction. 
As the current density must follow this direction at the entry and exit points of the molecule,
and as the momentum is very large at such high bias voltages,
reflection at the ZGNR-edges most probably leads to the meandering course observed in the current dynamics.

Regarding the effect of the central group,
it can be observed that the influence of the electron withdrawing increases at larger bias voltages.
This can be due to the availability of a larger number of charge carriers, that react to the induction of
the nitro group.
For the negative bias voltage ramp (cf. Fig.~\ref{fig:je_4ps}b), the electron dynamics proceeds via its presumably preferred pathway on the right edge of the central group.
The nitro group enhances this effect by concentrating the current density on one side of the phenyl group.
For the positive bias voltage ramp (cf. Fig.~\ref{fig:je_4ps}a), the nitro group has the opposite effect and drags
current density onto the other side of the ring.
In contrast to the  previous example at lower bias (cf. Fig.~\ref{fig:je_05Vb}),
this effect is strong enough to divert most of the current density to that side of the ring.
This shows that, despite the two configurations being energetically equivalent,
the nanojunction will exhibit a slight asymmetry upon reversal of the current direction.

\section{Conclusions}

Nitro-substituted oligo(phenylene-ethynylene) covalently bound between two ZGNR electrodes is a molecular junction 
with great potential for nanoelectronic applications.
Recently, we demonstrated using parameter-free quantum dynamical  modeling that 
this system can be switched reliably and reversibly between two logical conformers -- a planar conducting (ON) and a perpendicular less conducting conformer (OFF) -- by application of a gate electric field.
In the present work, we applied the driven Liouville-von-Neumann (DLvN) approach for time-dependent electronic
transport calculations to investigate the electronic current dynamics in this nanojunction at different applied bias voltages.
To this end, we introduced a partitioning procedure for the Hamiltonian based on the localization of orthogonal molecular
orbitals obtained from a standard ground state  density functional theory calculation.
Here, we could show that although the resulting energy spectra are nearly identical for the ON and OFF conformers,
they exhibit widely different conduction properties.
This was confirmed in the subsequent time-dependent analysis, where a linear voltage ramp was applied and the current passing through the device was monitored.
While in the OFF position the current through the junction always stays negligibly small, 
in the ON state, different transport channels are successively opened as the bias is increased.
This leads to a series of distinct steps in the current voltage characteristics.
In order to achieve convergence with respect to non-equilibrium Green's function (NEGF) reference simulations,
the system was extended by additional lead units using a microscopically parametrized tight binding Hamiltonian.
Introducing additional units as buffer units between the lead and the extended molecule
diminished further artificial coupling between the implicit electron reservoir and the extended molecule and lead to a quantitative convergence.

The mechanistic details of the charge transfer were investigated using the electronic current density,
which describes the spatially resolved instantaneous flow of electrons.
The first major focus of this work was the ultrafast equilibration dynamics of the incoming electronic current density, when a small bias voltage is suddenly applied.
Here, it was found that the incoming electron flow exhibits typical hydrodynamic properties, 
where the electrons propagate through the $\pi$-system mainly along the bonds, 
and with moderate influence of the molecular structure on ultrafast timescales.
Recently, similar hydrodynamic behavior of the electronic flow has been demonstrated experimentally.\cite{Torre2015electronViscosity,Levitov2016electronViscosity,Bandurin2016electronViscosity}

Applying a time-dependent linear bias voltage ramp to the junction,
the current density was found to follow a laminar course along the edge for small and a meandering course
at large bias voltages.
This curved path is caused by reflections of the current density at the ZGNR edges due to the large momentum combined with the angle between the overall direction of the electron transport and the transport axis defined by
the central group.
Consequently, bringing both axes into maximum coincidence could lead to an enhancement of the conductivity
of the junction.
This could be achieved, e.g., by choosing the same lattice direction which would lead to an armchair GNR (AGNR), or by preserving the lattice direction of the ZGNR contacts and using pyrrole rings to connect the leads with the central switching unit.
This could potentially lead to an enhancement of the conductivity of the junction.
Note that especially narrow AGNR are often semiconductors, and thus, not suitable as lead material.
In summary, we believe that the new imaging tool presented in this work
-- the electronic current density --
could potentially become very useful for understanding the electron transport in molecular junctions.

\section*{Acknowledgements}
The authors gratefully acknowledge the Scientific Computing
Services Unit of the Zentraleinrichtung f\"ur Datenverarbeitung (Zedat)
at Freie Universt\"at Berlin for allocation of computer time.
Furthermore, we thank Hans-Christian Hege for providing the ZIBAmira visualization program. 
The funding of the Deutsche Forschungsgemeinschaft (project TR1109/2-1 and Priority Program (SPP) 1459), 
from the Studienstiftung des deutschen Volkes e.V., 
and from the Elsa-Neumann foundation of the Land Berlin is also acknowledged.
We further thank the International Max Planck Research School "Complex Surfaces in Material Sciences" for its support.

\providecommand*{\mcitethebibliography}{\thebibliography}
\csname @ifundefined\endcsname{endmcitethebibliography}
{\let\endmcitethebibliography\endthebibliography}{}

\clearpage


\section*{Supplementary material}

# Mechanistic Investigations of Electronic Current Dynamics Through a Single-Molecule-Graphene–Nanoribbon Junction

Vincent Pohl[*] and Lukas Eugen Marsoner Steinkasserer

*Institute for Chemistry and Biochemistry, Freie Universität Berlin, Takustraße 3, 14195 Berlin, Germany*

Jean Christophe Tremblay[†]

*Institute for Chemistry and Biochemistry, Freie Universität Berlin, Takustraße 3, 14195 Berlin, Germany and*
*Laboratoire de Physique et Chimie Théoriques CNRS-Université de Lorraine,*
*UMR 7019, ICPM, 1 Bd Arago, 57070 Metz, France*

### Appendix A: Time Evolution of the Pseudo-Spectral State Populations

To confirm that the Pauli principle is satisfied by the density matrix obtained from the driven Liouville-von Neumann simulations, we can monitor the population of the pseudo-spectral states used in the basis. As a rule of thumb, the Pauli principle can be said to be satisfied provided all populations always lie between zero and one at all times. As can be seen from Fig. S1, this condition is indeed respected throughout our simulation, also at large biases. As the bias voltage increases, new transport channels are subsequently populated and depopulated. The subsequent relaxation dynamics towards population equilibration can take up to picoseconds. In the right panel, more channels are present due to the tight-binding extension but the general behaviour remains the same and the Pauli principle remains fulfilled.

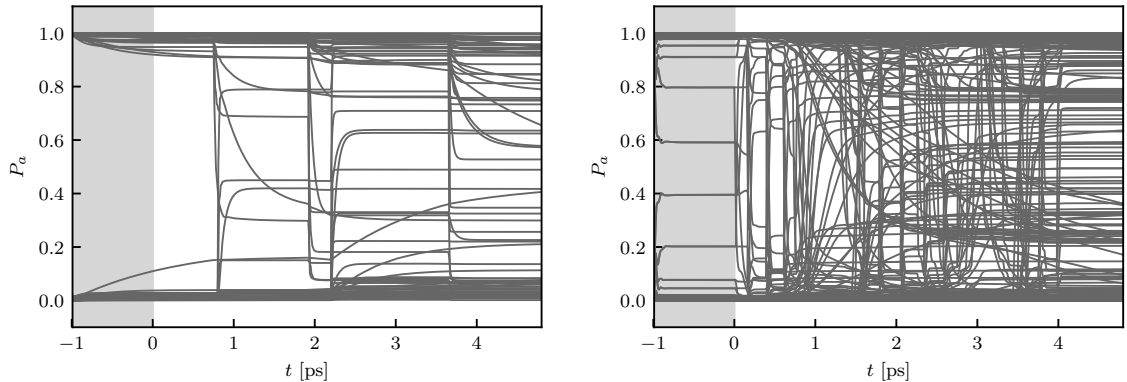

FIG. S1: Time-dependent state populations of the OPE-GNR junction under a linear voltage ramp from $U(t=0\,\mathrm{ps}) = 0$ to $U(t=4\,\mathrm{ps}) = 4\,\mathrm{V}$ for the ON conformation (cf. Fig. 5). Left panel: The time-dependent populations for the system without tight-binding extension. Right panel: The time-dependent populations for a system extended using the tight-binding Hamiltonian with ten buffer and ten lead units.

---

[*] v.pohl@fu-berlin.de
[†] jean-christophe.tremblay@univ-lorraine.fr

## Appendix B: Convergence of the Electronic Continuity Equation

In this section, we present two perspectives to estimate the convergence of the electronic continuity equation in position space: from the time-evolution of the block of the density matrix corresponding to the extended molecule (cf. Eq. 18) in position representation

$$\frac{\partial \langle \vec{r} | \underline{\underline{\rho}}_\mathrm{M}(t) | \vec{r} \rangle}{\partial t} = \underbrace{-\frac{\imath}{\hbar} \langle \vec{r} | \left[ \underline{\underline{H}}_\mathrm{M}, \underline{\underline{\rho}}_\mathrm{M}(t) \right] | \vec{r} \rangle}_{+\widetilde{\varrho}_\mathrm{M}(\vec{r},t)} \qquad (\mathrm{B1})$$

$$\underbrace{-\frac{\imath}{\hbar} \langle \vec{r} | \left( \underline{\underline{V}}_\mathrm{ML} \underline{\underline{\rho}}_\mathrm{LM}(t) - \underline{\underline{\rho}}_\mathrm{ML}(t) \underline{\underline{V}}_\mathrm{LM} \right) | \vec{r} \rangle}_{+\widetilde{\varrho}_\mathrm{LM}(\vec{r},t)} \quad \underbrace{-\frac{\imath}{\hbar} \langle \vec{r} | \left( \underline{\underline{V}}_\mathrm{MR} \underline{\underline{\rho}}_\mathrm{RM}(t) - \underline{\underline{\rho}}_\mathrm{MR}(t) \underline{\underline{V}}_\mathrm{RM} \right) | \vec{r} \rangle}_{+\widetilde{\varrho}_\mathrm{RM}(\vec{r},t)} ,$$

or from the position representation of the full driven Liouville-von Neumann equation

$$\frac{\partial \langle \vec{r} | \underline{\underline{\hat{\rho}}}(t) | \vec{r} \rangle}{\partial t} = \underbrace{-\frac{\imath}{\hbar} \langle \vec{r} | \left[ \hat{H}, \underline{\underline{\hat{\rho}}}(t) \right] | \vec{r} \rangle}_{-\vec{\nabla}_e \cdot \vec{j}_\mathrm{M}(\vec{r},t)} \quad \underbrace{-\frac{\imath}{\hbar} \langle \vec{r} | \left[ \imath \hat{W}, \underline{\underline{\hat{\rho}}}(t) \right]_+ | \vec{r} \rangle}_{\mathcal{W}(\vec{r},t)} . \qquad (\mathrm{B2})$$

The right hand sides of both equations each consist of two parts: the coherent (blue braces) and the driving part (green braces). In Eq. (B1), the electron flow within the extended molecule $\widetilde{\varrho}_\mathrm{M}(\vec{r},t)$ describes the coherent electronic contribution, and the incoherent flow of electrons from the leads to the central unit is given by $\widetilde{\varrho}_\mathrm{LM}(\vec{r},t) + \widetilde{\varrho}_\mathrm{RM}(\vec{r},t)$. In Eq. (B2), the spatial divergence of the flux density $-\vec{\nabla}_e \cdot \vec{j}_\mathrm{M}(\vec{r},t)$ is used to describe the coherent electron flow, and the complex Hamiltonian $\mathcal{W}(\vec{r},t)$ yields the incoherent contribution to the electronic current, which we dub "driving term".

Integrating the above quantities over a volume $\Omega_y$ with proper boundary conditions,

$$\begin{array}{lll} x_\mathrm{min} = -\infty & y_\mathrm{min} = -\infty & z_\mathrm{min} = -\infty \\ x_\mathrm{max} = +\infty & y_\mathrm{max} = y & z_\mathrm{max} = +\infty \end{array} . \qquad (\mathrm{B3})$$

and multiplying the results by the elementary charge, yields relations for the charge conservation along the transport axis for each value of $y \in \mathrm{M}$ on the extended molecule. For a quasi-stationary state, the coherent and the driving term have opposite signs and should have the same absolute value, i.e.,

$$\begin{aligned} I(t) &= e \iiint_{\Omega_y} \widetilde{\varrho}_\mathrm{M}(\vec{r},t) \mathrm{d}^3 \Omega_y & &= -e \iiint_{\Omega_y} \left( \widetilde{\varrho}_\mathrm{LM}(\vec{r},t) + \widetilde{\varrho}_\mathrm{RM}(\vec{r},t) \right) \mathrm{d}^3 \Omega_y \\ &= e \iiint_{\Omega_y} \left( -\vec{\nabla}_e \cdot \vec{j}_\mathrm{M}(\vec{r},t) \right) \mathrm{d}^3 \Omega_y & &= -e \iiint_{\Omega_y} \mathcal{W}(\vec{r},t) \mathrm{d}^3 \Omega_y \\ &= e \oiint_{S_y} \left( -\vec{j}_\mathrm{M}(\vec{r},t) \cdot \vec{n}_y \right) \mathrm{d}^2 S(y). & & \end{aligned} \qquad (\mathrm{B4})$$

The last line is a consequence of the divergence theorem for vanishing current far away from the molecular device in the direction perpendicular to the electron flow. The Eqs. (B4) can be used to evaluate the convergence of continuity equation with respect to the basis set, the grid, and the method in general.

Fig. S2 shows these contributions for a representative snapshot for $U = 4.8\,\mathrm{V}$ in the ON conformation. All contributions have the same general behavior: the absolute value rises until the boundaries of the central molecule ($y \approx -20\,\text{Å}$) to a plateau with the height of $I(t)$ (black dashed dotted line, cf. Eq. (20)). The value stays constant within the spatial extent of the extended molecule, until it finally decreases again for $y \gtrsim 20\,\text{Å}$. The deviations from the $I(t)$ reference computed using Eq. (20) are smaller for the contributions derived from the density than those derived from the flux density (blue and orange curve in the right panel of Fig. S2). In particular, $e \iiint_{\Omega_y} \left( -\vec{\nabla}_e \cdot \vec{j}_\mathrm{M}(\vec{r},t) \right) \mathrm{d}^3 \Omega_y$

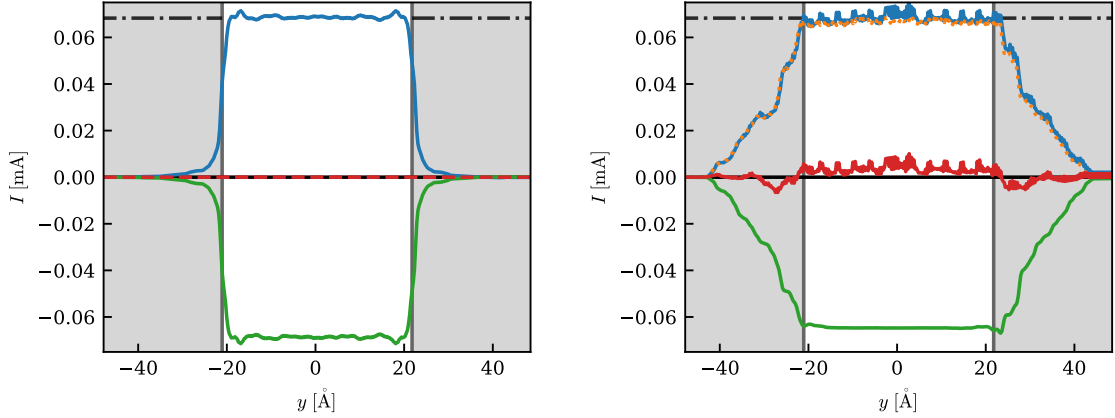

FIG. S2: The quantities defined in Eg. (B4) exemplary shown for a representative snapshot at $t = 4.8\,\mathrm{ps}$ for a simulation with a linear voltage ramp from $U(t = 0\,\mathrm{ps}) = 0$ to $U(t = 4\,\mathrm{ps}) = 4\,\mathrm{V}$ for the ON conformation. The simulation has been performed without any tight-binding extension at the PBE/def2-SVP level of theory. Energy windows of $\Delta E_{\mathrm{lead}} = 6\,\mathrm{eV}$, $\Delta E_{\mathrm{basis}} = 8\,\mathrm{eV}$ were used for the localization. The black dashed dotted line is the reference value corresponding to the current definition as given in Eq. (20). The black solid line corresponds to the integral over $\frac{\partial}{\partial t}\left\langle \vec{r}\left|\hat{\hat{\rho}}(t)\right|\vec{r}\right\rangle$. Left panel: Quantities derived from Eq. (B1). Blue curve: the current within the extended molecule $e\iiint_{\Omega_y}\widetilde{\varrho}_{\mathrm{M}}(\vec{r},t)\mathrm{d}^3\Omega_y$, and green curve: the current from the leads to the central unit $e\iiint_{\Omega_y}\bigl(\widetilde{\varrho}_{\mathrm{LM}}(\vec{r},t)+\widetilde{\varrho}_{\mathrm{RM}}(\vec{r},t)\bigr)\mathrm{d}^3\Omega_y$. The red dashed curve correponds to the sum of both. Right panel: Quantities derived from Eq. (B2). Blue solid curve: the integral over the negative spatial divergence of the current density $e\iiint_{\Omega_y}\bigl(-\vec{\nabla}_e\cdot\vec{j}_{\mathrm{M}}(\vec{r},t)\bigr)\mathrm{d}^3\Omega_y$, orange dotted curve: the surface integral over the current density $e\oiint_{S_y}\bigl(-\vec{j}_{\mathrm{M}}(\vec{r},t)\cdot\vec{n}_y\bigr)\mathrm{d}^2S(y)$ and green curve: the integral over the position representation of the complex Hamiltonian $e\iiint_{\Omega_y}\mathcal{W}(\vec{r},t)\mathrm{d}^3\Omega_y$. The red curve corresponds to the sums of the blue and the green curves.

shows small oscillations, which can be attributed to the inaccurate representation of the cusps on the atoms when using atom-centered Gaussian functions as a primitive basis set. This limitation was described in detail in Ref. [1].

The simulation has been performed without any tight-binding extension at the PBE/def2-SVP level of theory. Energy windows of $\Delta E_{\mathrm{lead}} = 6\,\mathrm{eV}$, $\Delta E_{\mathrm{basis}} = 8\,\mathrm{eV}$ were used for the localization. Note that tight-binding extensions within this work are done in the energy space, and, consequently, no spatial information is available on the leads nor in the buffer region.

### References

[1] V. Pohl, G. Hermann and J. C. Tremblay, *J. Comput. Chem.*, 2017, **38**, 1515–1527.



\end{document}